# Quantum Information Aspects on Bulk and Nano Interacting Fermi System: Spin-Space Density Matrix Approach

R. Afzali[1], N. Ebrahimian[2], B. Eghbalifar[3]

[1]Department of Physics, K. N. Toosi University of Technology, Tehran, 15418, Iran,

[2]Department of physics, Faculty of Basic Sciences, Shahed University, Tehran, 18155-159, Iran

[3]Yasouj branch, Azad University, Yasouj, Iran.

Emails: afzali@kntu.ac.ir, n.ebrahimian@shahed.ac.ir , b.eghbali2011@yahoo.com

## Abstract

In this paper, we investigate quantum correlation of an interacting Fermi system, which is a nodal superconductor (d-wave superconductor) at zero temperature, via quantum entanglement of two electron spins forming Cooper pairs (Werner state), tripartite and quantum discord. The energy gap depends on the angle between the electron momentum and the nodal axis; and at zero temperature we use an approximation in which the energy gap is considered as the linear function of the angle. After calculating single-electron Green's functions, the two-electron space-spin density matrix, which has X-state form, is obtained. The dependence of quantum correlation to the relative distance of electrons spins of Cooper pair and energy gap is investigated. One of the results is, for d-wave case, concurrence (as a measure of entanglement), quantum discord and tripartite are sensitive to the change of magnitude of gap. Another result is both concurrence and discord oscillate. Then, we consider three-dimensional rectangular nano-superconducting grain in the weak coupling frame. The nano-size effect is entered via gap fluctuation. The dependence of quantum correlation to length of superconductor and lower bound of robustness of tripartite entanglement are determined. Moreover, we show that quantum correlation of d-wave nano-size superconducting grain strongly depends on length of grain (in contrast to s-wave case). In general, it is found that the length of grain lower, the effect of nano-size on quantum correlation higher. Quantum tripartite for nano-scale d-wave superconductor is better than for bulk d-wave superconductor. However, we find out both bulk and nano-size s-wave superconductors have the same tripartite. Furthermore, entanglement length and quantum correlation length are investigated and it is shown that there is a length of superconductor in which discord becomes zero. Also, for a given fixed length of superconductor, both a peak in discord and a peak in concurrence occur simultaneously.

## 1. Introduction

Entanglement and quantum discord (QD) are the key resources in quantum communication, quantum teleportation and quantum computation. Quantum entanglement (QE) is a physical resource, like energy, associated with quantum correlations that are possible between separated quantum systems [1-10]. One of the measure of quantum entanglement namely concurrence can be used for the determination of correlation of systems [11-14]. A study of QD and QE in many-body systems are very important to give new insights on physical properties via correlations, however, QD and QE have many applications to quantum information processing and to protocols such as quantum teleportation and quantum algorithm. On the other hand, QD and QE can be used to determine quantum phase transitions [15-18]. QD is defined as the difference between quantum mutual information and classical correlation in a bipartite system. In general, quantum discord may be nonzero even for certain separable states namely when entanglement of system is zero. QD can be considered as resource for remote state preparation [19]. QD specifies the interferometric power of quantum states [20].

Entanglement in many-body systems was studied [21]. The history of the investigation entanglement of spins returns to Ref. [16] that the properties of entangled systems in the second quantization formalism were studied; one of the results of this study is that at zero temperature a non-interacting Fermi system can be entangled in spin, providing that the distances of particles or quasiparticles do not exceed the inverse Fermi wavenumber. In Ref. [22], the entanglement of electron spins of non-interacting electron gases based on the Green's function approach was discussed. By considering the screened Coulomb interaction between electrons, the entanglement between two electrons in a degenerate electron gas was studied and it was shown that the interaction leads to a suppression of the entanglement distance [23] and the temperature was considered in Ref. [24].

The investigation bipartite entanglement (BE) in the s-wave superconductor was already done [25] and entanglement of two electron spins forming Cooper pairs was investigated, using two-electron space-spin density matrix. This two spin state is Werner state [25]. Also, in momentum space, the expression of the spin entanglement of electrons in Cooper pairs was driven [26] and the brief discussion of BE on FFLO superconductor was already done [27]. Furthermore, for finite superconductor, using the average local concurrence, entanglement of the full system was discussed [28].

Quantum tripartite and multipartite entanglement in a non-interacting fermion gas in terms of fermion separation were investigated, using density matrix [29]. It was proven that multipartite entanglement can be established only out of two-fermion entanglement. Dependence of tripartite entanglement (TE) of electron spins of a non-interacting electron gases with respect to the relative distance between the three spins and the temperature was determined [30]. Also, at zero temperature, in the non-interacting Fermi gas, tripartite shared among of three fermions was investigated [31] and for a given special configuration, using entanglement witnesses, it is shown that the three fermions are tripartite entangled. Furthermore, it is found that TE does not exist below a limitation.

QD was defined as a measure of the quantumness of correlations [32]. Necessary and sufficient condition for nonzero quantum discord was driven [33]. Review of quantum discord in bipartite and multipartite systems was done [34]. Quantum discord for two-qubit systems was driven [35,36] and for two-qubit X-states was calculated [37,38].

In condensed matter physics, it was known that d-wave symmetry of superconductor is more important than s-wave symmetry which supposed energy gap is constant. D-wave superconductors are considered as unconventional superconductors. Usually high temperature superconductors (HTSC) have the gap with d-wave symmetry (of course HTSC usually are with strong coupling regime). It is important that one get the knowledge of a superconductor with momentum-dependence gap. Therefore, we pay attention on d-wave case. For obtaining the concurrence and bipartite entanglement of electron spins, we must calculate the two-electron space-spin density matrix with the aid of two-particle Green's function. Meanwhile, for some purposes we calculate an analytic form of dominated Green's function, but for studying effect of length of d-wave nano-superconductor grain or when concentrating on the gap change, the numerical calculation of Green's function is used. Then, the two-electron space-spin density matrix can be written in terms of normal and anomalous single-particle Green's functions. For investigating TE of electron spins of Cooper pair of d-wave superconductor as an interacting system, new parameters for three-spin reduced density matrix is calculated. Robustness of TE, defined in Refs. [31,39-40], is obtained. Lower bound of robustness of TE, $E_{R,\min}\left(\circ_3\right)$, is determined and the role of interaction that principally revealed via gap, whether in d-wave or in s-wave, is identified. Then, quantum discord is presented. Finally, we suppose that the single particle level spacing of system is much smaller than energy gap and this assumption guarantee to satisfy BCS model [41]. It is seen that in general nano-size effect, which entered via gap fluctuation (thereby there is the change in the interaction), influence deeply and widely on quantum correlations. Therefore, nano-size effect is more efficient on properties of system via the change of quantum correlation. Dependence of correlation to electrons distance and length of the superconductor in spin space is determined.

Paper will be organized as follows. In section 2, after writing the Hamiltonian of d-wave superconductor and energy gap function, in order to calculate the quantum entanglement, we obtain Green's functions and thereby we get density matrix of system. In section 3,4 and 5, we bring the results of the concurrence, the calculation of the three-spin reduced density matrix accompanied with the identification of TE and $E_{R,\min}$ , and the behavior of QD, respectively. In section 6, the nano-size effect on quantum correlation is determined and discussed in details.

## 2.Quantum Entanglement

First of all, we proceed to obtain quantum bipartite entanglement of system. For this purpose, we calculate concurrence of system. Therefore, we start to obtain density matrix of system by using Green's function of d-wave superconductor.

Hamiltonian of d-wave superconductor is given by [42]

$$H=\sum_{ks}\varepsilon_k c_{ks}^{\dagger}c_{ks}+\sum_{k,k'}V_{k,k'}c_{k\uparrow}^{\dagger}c_{-k\downarrow}^{\dagger}c_{-k'\downarrow}c_{k'\uparrow} \qquad (1)$$

where $\varepsilon_k$ , $c_{ks}^{\dagger}$ and $c_{ks}$ are the excitation energy with respect to chemical potential, the creation and the annihilation operators, respectively. The interaction potential of d-wave superconductor, $V_{k,k'}$ , is given by[43,44]

$$V_{kk'}(\vec{v}_F,\vec{v}'_F)=V_d\cos2(\theta_k-\alpha)\cos2(\theta'_k-\alpha) \qquad (2)$$

where $\vec{v}_F$ , $\alpha$ and $\theta_k$ (and $\theta'_k$) are Fermi velocity, the angle between crystallographic a-direction and the x-axis, and the direction of $\vec{v}_F$ ($\vec{v}'_F$) in the ab-plane. Also, $V_d$ is defined via the dimensionless BCS constant of interaction $\lambda_d$ that given by $\lambda_d=V_d N(0)/2$ wherein $N(0)$ is density of states. On the other hand, Hamiltonian can be written in terms of gap energy using mean field approximation as

$$H=\sum_{ks}\varepsilon_k c_{ks}^{\dagger}c_{ks}+\sum_{k,k'}\Delta_k c_{k\uparrow}^{\dagger}c_{-k\downarrow}^{\dagger}+\sum_{k,k'}\Delta^{*}{}_k c_{-k'\downarrow}c_{k'\uparrow} \qquad (3)$$

Indices $k$ and $s$ denote wave vector and spin component, respectively. For d-wave case with $d_{x^2-y^2}$ symmetry, we have

$$\Delta_k = \Delta\left(\hat{k}_x^2 - \hat{k}_y^2\right) = \Delta \cos\left(2\,\bar{}_{k}\right) \quad (4)$$

where $\bar{}_{k}$ is the angle between electron momentum and gap axis and $\Delta$ is magnitude of gap. D-wave gap goes to zero on the Fermi surface at four nodes, where low-energy excitations are possible. Order parameter with d-wave symmetry has four nodal points, $\vec{k}_i = \left(\pm k_F/\sqrt{2}\ \text{و}\ \pm k_F/\sqrt{2}\right)$ , where the Fermi surface crosses the nodal directions $k_x = \pm k_y$ . In these points, $\Delta_k$ turns zero. Furthermore, we can approximate the energy gap at low temperatures [45]. For this limit, and for d-wave case, $\Delta_k$ is $2\Delta_0\ddot{}$ where $\ddot{}$ is the angular deviation of $\hat{k}$ and $\Delta_0$ is the maximum gap function.

Now we proceed to calculate density matrix of system. Density matrix of system is defined by [16,25]

$$\circ^{(2)}\left(x_1, x_2; x_1', x_2'\right) = (1/2)\left\langle \hat{\Psi}^{\dagger}\left(x_2'\right)\hat{\Psi}^{\dagger}\left(x_1'\right)\hat{\Psi}\left(x_1\right)\hat{\Psi}\left(x_2\right)\right\rangle \quad (5)$$

where $\hat{\Psi}^{\dagger}(x)$ and $\hat{\Psi}(x)$ are creation and annihilation field operators of particles, respectively, and $\langle\ \rangle$ means $\langle \cdot_0 | \ | \cdot_0 \rangle$ with BCS ground state $\cdot_0$ and $x \equiv (\vec{x}, s)$ and $s$, for example, can be considered as spin. It merits mentioning that field operators satisfy anticommutation relations, because of Fermi statistics [46]. Density matrix in terms of two electron spin-space Green's function can be written as

$$\circ^{(2)}\left(x_1, x_2, x_1', x_2'\right) = -(1/2)\,G\left(x_1 t_1, x_2 t_2, x_1' t_1^{+}, x_2' t_2^{+}\right) \quad (6)$$

where $t_1^{+}$ means $t_1 + 0^{+}$ and Green's function is defined by

$$G\left(1, 2, 1', 2'\right) = -\left\langle T\left[\hat{\Psi}_H(1)\hat{\Psi}_H(2)\hat{\Psi}_H^{\dagger}(2')\hat{\Psi}_H^{\dagger}(1')\right]\right\rangle \quad (7)$$

$T$ is time ordering, subscript $H$ denotes as Heisenberg picture and $(i) \equiv (x_i, t_i)$. The two electron spin-space Green's function is related to single-particle Green's function via the following equation [25,46]

$$G\left(1, 2, 1', 2'\right) = G\left(1, 1'\right)G\left(2, 2'\right) - G\left(1, 2'\right)G\left(2, 1'\right) - F\left(1, 2\right)F^{\dagger}\left(1', 2'\right) \quad (8)$$

where single-particle Green's function and anomalous Green's function are defined as [25,46]

$$G(1,1') = -i\left\langle T\left[\Psi_H(x_1t_1)\Psi^{\dagger}_H(x_1't_1')\right]\right\rangle = \delta_{s_1s_2'}G\ (r_1t_1, r_1't_1') \tag{9}$$

and

$$F^{\dagger}(1,2) = -i\left\langle T\left[\Psi^{\dagger}_H(x_1t_1)\Psi^{\dagger}_H(x_2t_2)\right]\right\rangle = I_{s_1s_2}F^{\dagger}(r_1t_1, r_2t_2) \tag{10}$$

where $I_{s_1s_2} \equiv i\sigma_y$ [25]. One of the property of $F$ (and $F^{\dagger}$) is

$$F(\vec{x}_1,\vec{x}_2) = F(\vec{x}_2,\vec{x}_1)\ ,\quad F^{\dagger}(\vec{x}_1,\vec{x}_2) = F^{\dagger}(\vec{x}_2,\vec{x}_1) \tag{11}$$

It should be noted that the spin dependence of both normal and anomalous Green's function are given by $\delta_{s_1s_2'}$ and $I_{s_1s_2}$, respectively. By using Eqs. (7)-(10), Eq. (6) becomes

$$\begin{aligned}\Gamma^{(2)}_{s_1,s_2;s_1',s_2'}(r_1,r_2;r_1',r_2') = -(1/2)\big[&\delta_{s_1s_1'}\delta_{s_2s_2'}G(r_1-r_1')G(r_2-r_2')\\ &-\delta_{s_1s_2'}\delta_{s_2s_1'}G(r_1-r_2')G(r_2-r_1')\\ &-I_{s_1s_2}I_{s_1',s_2'}F(r_1-r_2)F^{*}(r_1'-r_2')\big]\end{aligned} \tag{12}$$

We will see that in contradiction to s-wave case, Green's functions for d-wave case, are not pure imaginary but instead Green's functions are complex and therefore, Green's functions do not depend on only magnitude of relative distance of two electrons. Now for obtaining $\Gamma^{(2)}$, we proceed to calculate single-particle Green's function. For d-wave superconductor, we have

$$G = \frac{-1}{\omega_n^2+\xi_k^2+\Delta_k^2}\begin{bmatrix} i\omega_n+\xi & \Delta_k \\ \Delta_k & i\omega_n-\xi_k\end{bmatrix} \equiv \begin{bmatrix} G_{11} & G_{12} \\ G_{21} & G_{22}\end{bmatrix} \tag{13}$$

On the other hand, we have

$$G_{11}(k,i\omega_n) \equiv G = -\frac{i\omega_n+\xi_k}{\omega_n^2+\xi_k^2+\Delta_k^2}\ ,\quad G_{22}(k,i\omega_n) \equiv -G^{*} \tag{14}$$

$$G_{12}(k,i\omega_n) \equiv F = -\frac{\Delta_k}{\omega_n^2+\xi_k^2+\Delta_k^2}\ ,\quad G_{21}(k,i\omega_n) \equiv F^{\dagger} \tag{15}$$

where $\omega_n$ is Matsubara frequency and is given by $\omega_n = T(2n+1)\pi$ (throughout the paper $K_B = \hbar = 1$ is used) and $T$ is temperature. By using the transformation $\omega_n \to -i\omega$ , we can provide the formula for the limit $T \to 0$ . Then, we can write the Green's functions as follows

$$G_{11}(k,\omega) \equiv G = \frac{\omega + \xi_k}{\omega^2 - \xi_k^2 - \Delta_k^2} \ , \tag{16}$$

$$G_{12}(k,\omega) \equiv F = -\frac{\Delta_k}{\omega^2 - \xi_k^2 - \Delta_k^2} \tag{17}$$

Now we need to have the single-particle Green's function in coordinate-time space. Therefore, first of all, we provide Fourier transformation from $\omega$ to time space. We have

$$G_{11}(k,t) \equiv \lim_{t\to 0} \int_{-\infty}^{+\infty} e^{i\omega t} \frac{\omega + \xi_k}{\omega^2 - \xi_k^2 - \Delta_k^2} d\omega = -i\pi \left( 1 - \frac{\xi_k}{\sqrt{\xi_k^2 + \Delta_k^2}} \right) \tag{18}$$

and

$$G_{12}(k,t) \equiv \lim_{t\to 0} \int_{-\infty}^{+\infty} e^{i\omega t} \frac{-\Delta_k}{\omega^2 - \xi_k^2 - \Delta_k^2} d\omega = -i\pi \frac{\Delta_k}{\sqrt{\xi_k^2 + \Delta_k^2}} \tag{19}$$

Now we make Fourier transformation from $k$ to $r$ -space and use $\Delta_k = 2\Delta_0 \varphi$ [45]. we get

$$iG_{11}(r) \equiv \pi \int_{-\infty}^{+\infty} e^{ikr\cos\left(\varphi + \frac{\pi}{4}\right)} \left( 1 - \frac{\xi_k}{\sqrt{\xi_k^2 + 4\Delta_0^2 \varphi^2}} \right) d^2k \tag{20}$$

In fact, we have concentrated in quasi-two dimensional space thus we have written $d^2k\left(= k\ dk d\theta\right)$. Of course, the integration over $\theta$ can be replaced by $\varphi$ (the angle between momentum of electron and node axis) as [45]

$$\int_0^{\pi} d\theta = \int_{-\frac{\pi}{4}}^{\frac{\pi}{4}} d\varphi \quad , \quad \theta = \frac{\pi}{4} + \varphi \tag{21}$$

Also, we would like to replace the integration over energy instead of momentum. Therefore, we apply $\left(2®\right)^{-1} kdk = N\left(£\right)d£$ that is satisfied to the quasi-two dimensional space. Also, we can use the approximation $N\left(£\right) \simeq N\left(0\right)$ that is a good approximation at zero temperature [46]. Meanwhile, at low temperatures and zero temperature, the values of ¨ is located about the nodes and around zero value, therefore we can assume ¨ $\approx 0$ just at exponential existing in Eq. (20). Furthermore, we can use [25]

$$kr \simeq k_F r + yt \tag{22}$$

where

$$y = \frac{r}{®¶_0} \quad \text{and} \quad t = \frac{£}{\Delta_0} \tag{23}$$

wherein $¶_0$ is coherence length. Then, after integrating over the angle ¨ , we get

$$iG_{11}\left(r\right) \equiv \frac{\Delta_0 N\left(0\right)}{2} \mathrm{e}^{i\frac{k_F r}{\sqrt{2}}} \int dte^{iy't} \left(\frac{®}{2} - t\sinh\left(\frac{®}{2t}\right)\right) \tag{24}$$

where $y' = y/\sqrt{2}$ . It should be noted that the interval of integration is $-\mu_D/\Delta_0 \le t\left(\equiv £/\Delta_0\right) \le \mu_D/\Delta_0$ . Now, we consider the interval of integration as $-\infty \le t\left(\equiv £/\Delta_0\right) \le \infty$ to obtain analytical expression of Green's functions; we can use this approximation in some purposes. However, it should be noted that when we deal with to enter, for example, nano-size effect on correlation, we do not use this approximation. Furthermore, when we use this approximation, some errors occur; for example, as will seen, it causes that entanglement length becomes constant and also the oscillation of concurrence disappears. Nevertheless, after using the interval $-\infty \le t\left(\equiv £/\Delta_0\right) \le \infty$ and using Fourier transformation of Eq. (24), we have

$$\begin{aligned} iG_{11}\left(r\right) \equiv & -\frac{\sqrt{2}®^3}{8} ¶_0 \Delta_0 N\left(0\right) \frac{e^{i\frac{k_F r}{\sqrt{2}}}}{r} (2\,\mathrm{I}_0\left(\frac{r}{2\sqrt{2}¶_0}\right) - 2\,{}_1\mathrm{F}_2\left(\frac{1}{2};1,\frac{3}{2};\frac{r^2}{32¶_0{}^2}\right) \\ & + \frac{\mathrm{r}}{¶_0®\sqrt{2}} ({}_2\mathrm{F}_3\left(1,1;\frac{3}{2},\frac{3}{2},2;\frac{r^2}{32¶_0{}^2}\right)) - 2\,\mathrm{L}_0\left(\frac{r}{2\sqrt{2}¶_0}\right) - ¢\left(\frac{\mathrm{r}}{¶_0®\sqrt{2}}\right)\frac{4\,\mathrm{r}}{¶_0®\sqrt{2}}) \end{aligned} \tag{25}$$

where $I_n\left(z\right)$ , ${}_pF_q\left(a;b;z\right)$, $L_n\left(z\right)$ and $¢\left(z\right)$ are the modified Bessel function of the first kind, the generalized Hypergeometric function, the modified Struve function, and the Dirac delta function, respectively. Also, the anomalous Green's function for d-wave superconductor is negligible to comparison with normal Green's function (as explained for s-wave superconductor [25]). Now by using the expression of Green's function (Eq. (25)), we can obtain density matrix with aid of Eq. (12). On the other hand, we can express two-dimensional density matrix by [25]

$$^{\circ}_{12} = (1-p)\frac{I}{4} + p\left|\cdot^{(-)}\right\rangle\left\langle\cdot^{(-)}\right| \tag{26}$$

where $I$ is a $4\times4$ unit matrix and $\left|\cdot^{(-)}\right\rangle = \left(1/\sqrt{2}\right)\left(\left|\uparrow\downarrow\right\rangle - \left|\downarrow\uparrow\right\rangle\right)$ and $p$ is a parameter that identify a Werner state [47] and for d-wave case, the parameter is given by the following equation, which has a little different in comparison to s-wave case,

$$p = \frac{g_{11}(\vec{r})\,g_{11}(-\vec{r})}{2 - g_{11}(\vec{r})\,g_{11}(-\vec{r})} \tag{27}$$

where $g$ is defined by $g = G/G(0)$. For s-wave case and a non-interacting system, Eq. (27) convert to $\mathrm{P} = \left(g_{11}(\vec{r})\,g_{11}(\vec{r})\right)/\left(2 - g_{11}(\vec{r})\,g_{11}(\vec{r})\right)$. According to the Preres-Horodecki separability criterion, a Werner state is entangled for $p > 1/3$ [48]. The concurrence can be calculated by [49]

$$C = \max\left\{0, (3p-1)/2\right\} \tag{28}$$

By using Eqs. (24) and (27)-(28), the variations of concurrence in terms of the relative distance of two electrons of Cooper pairs ($r$) for various values of the energy gap magnitude (in all paper, the value of energy gap magnitude is measured with respect to Fermi energy that we have considered 1ev) is depicted in Fig. 1(a).

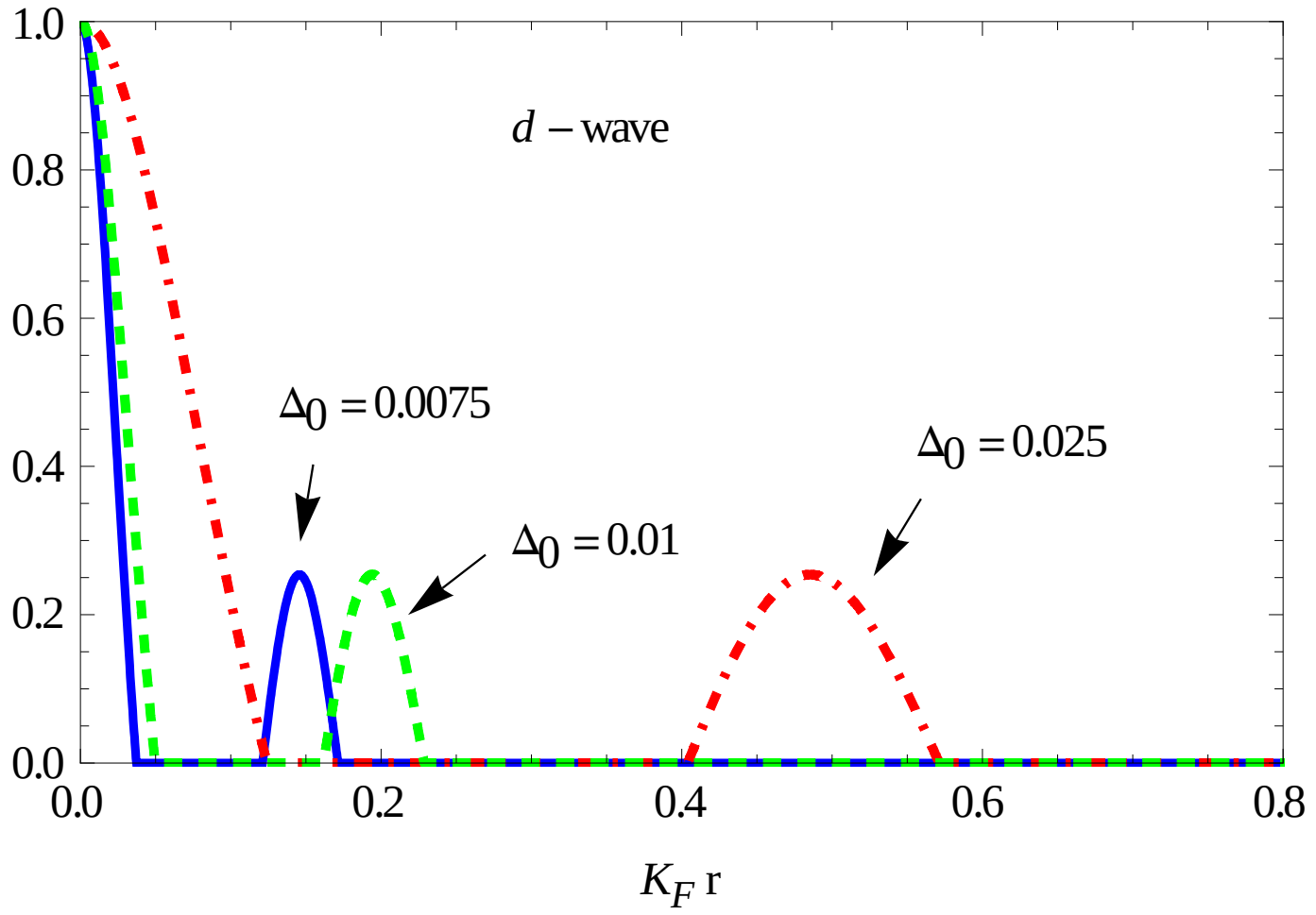

Fig.1(a): (Color online) For d-wave case; Concurrence versus relative distance of electrons of a Cooper pairs (times to $k_F$ ) for various values of the energy gap magnitude.

It is seen that at a fixed gap magnitude, in some interval of values $k_F r$ , which concurrence is nonzero, concurrence has a peak. For higher fixed gap magnitude, peak occurs at higher $k_F r$ (regardless of the first peak of concurrence at different gap magnitude as these peaks occur at the same $k_F r\,(=0)$ with the value one). In contrast to s-wave case, for d-wave case, concurrence (at fixed gap magnitude) oscillates. But for s-wave case, it was shown that the functional dependence of concurrence to $k_F r$ is just a decreasing function [25]. For d-wave case, at any different gap magnitude, there are two principal interval of values $k_F r$ in which concurrence is nonzero and in other intervals concurrence is zero (up to $k_F r = 100$, it was tested but we didn't bring in Fig. 1(a)).

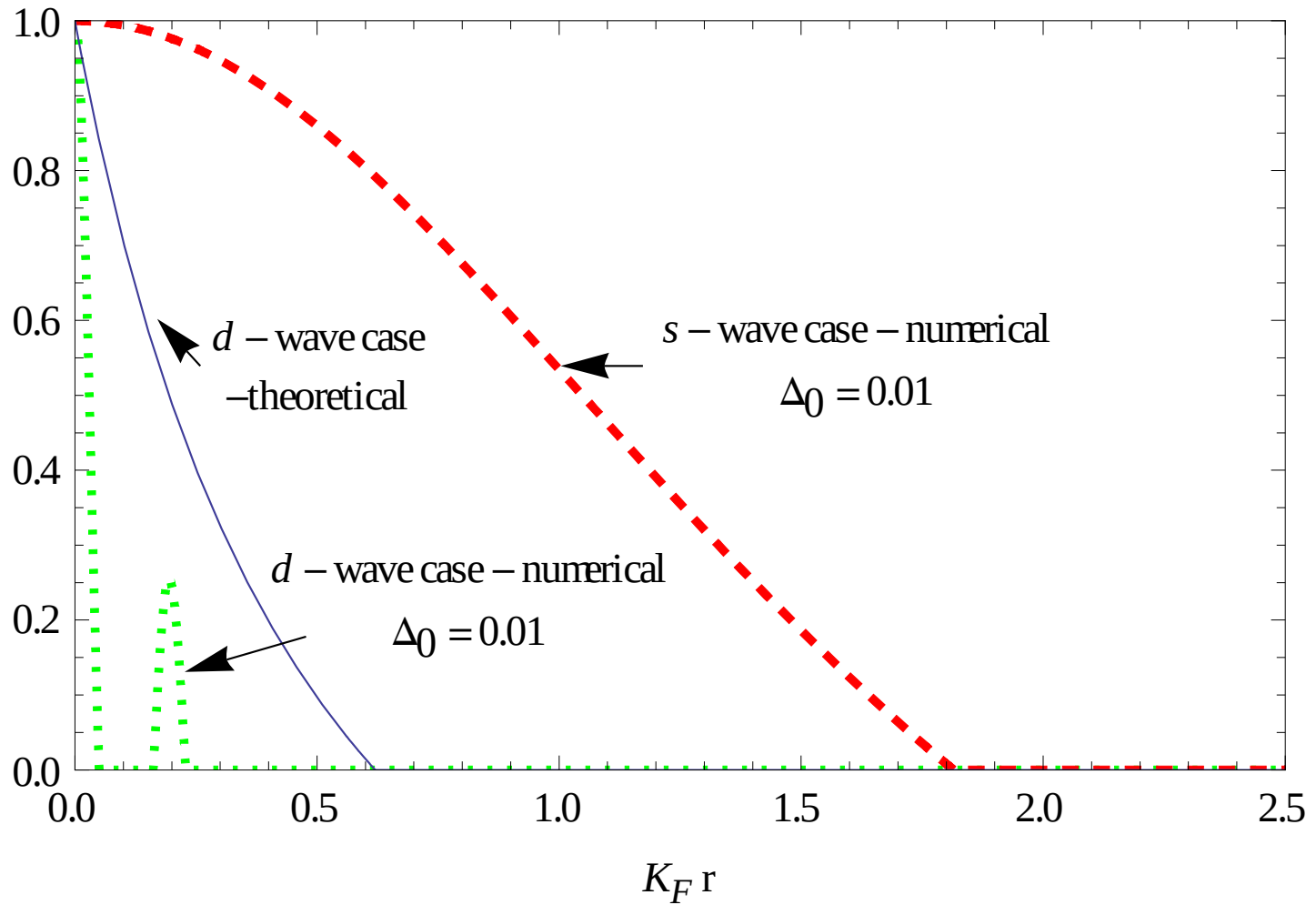

Fig.1(b): (Color online) Solid curve (dotted curve): Concurrence versus $k_F r$ using analytical (numerical) Green's function. Dashed curve: Concurrence for s-wave case. The value of gap magnitude was selected the same for d- and s- wave cases.

In Fig. 1 (b), for d-wave case, concurrence with using analytical Green's function (Solid curve) and also using numerical Green's function (dotted curve) were plotted. Also, we have brought concurrence for s-wave case (dashed curve) to compare with the d-wave case. As seen, entanglement length (for d-wave case, we would like to select the value of $k_F r$ that occurrence becomes zero at all namely after the ending of oscillation) for d-wave case is lower than for s-wave case. This result is correct for all the values of valid interval of gap magnitude; it merits mentioning that our work is in the weak coupling regime (strong coupling regime needs to applying Eliashberg equations [42] and we will apply this equations in future) where $\Delta_0 << \mu_D < \pounds_F$ is satisfied. Furthermore, in this valid regime, for s-wave case, entanglement length does not change by various gap magnitudes (just there is one curve as shown in Fig. 1 (b)) and also for d-wave case, for different gap magnitude located in this regime, entanglement length moves to higher $k_F r$ but takes lower value with respect to entanglement length related to s-wave case. It should be noted that in Fig. 1(b), concurrence of the s-wave case was plotted with different method of those was previously plotted by Kim et al.[25]. Of course, entanglement length, in which BE becomes zero, approximately reaches to the reported value i.e. $\approx 1.8/k_F$ for s-wave case. Also, it should be noted that the d-wave case, as a superconductor with nodes, is sensitive to different energy gap magnitude, therefore we used the numerical form of Green’s function. Furthermore, it is seem that a d-wave superconductor with very small value of gap magnitude can have very small entanglement length.

## 3. Tripartite entanglement

It is interested to consider the correlation of three fermions in a superconductor as an interacting system with nodal gap. In fact, two of fermion that contributed in the pair (Cooper pair) and the fermion that contributed as one of the electrons of the other Cooper pair must be considered to investigate tripartite quantum entanglement. In my opinion, at finite temperatures (which we do not discuss in this paper), the situation is much interesting, because of the existence of normal electrons (do not contribute in the Cooper pair) and super-electrons (contribute in the Cooper pair).

In this section, we proceed to calculate the quantum tripartite entanglement of d-wave superconductor. First of all, we must calculate Green’s function up to order 3 in terms of single-particle Green’s function.

$$
\begin{aligned}
&\left\langle \cdot^{\dagger}_{t''}(r'')\cdot^{\dagger}_{t'}(r')\cdot^{\dagger}_{t}(r)\cdot^{\dagger}_{s}(r)\cdot^{\dagger}_{s'}(r')\cdot^{\dagger}_{s''}(r'')\right\rangle = \\
&+\text{¢}_{t''s}\text{¢}_{t's'}\text{¢}_{ts''}G(\vec{r}-\vec{r}'')G(\vec{r}'-\vec{r}')G(\vec{r}''-\vec{r})-\text{¢}_{t''s}\text{¢}_{t's''}\text{¢}_{ts'}G(\vec{r}-\vec{r}'')G(\vec{r}''-\vec{r}')G(\vec{r}'-\vec{r}) \\
&-\text{¢}_{t''s'}\text{¢}_{t's}\text{¢}_{ts''}G(\vec{r}'-\vec{r}'')G(\vec{r}-\vec{r}')G(\vec{r}''-\vec{r})+\text{¢}_{t''s'}\text{¢}_{t's''}\text{¢}_{ts}G(\vec{r}'-\vec{r}'')G(\vec{r}''-\vec{r}')G(\vec{r}-\vec{r}) \\
&+\text{¢}_{t''s''}\text{¢}_{t's}\text{¢}_{ts'}G(\vec{r}''-\vec{r}'')G(\vec{r}-\vec{r}')G(\vec{r}'-\vec{r})-\text{¢}_{t''s''}\text{¢}_{t's'}\text{¢}_{ts}G(\vec{r}''-\vec{r}'')G(\vec{r}'-\vec{r}')G(\vec{r}-\vec{r}) \\
&-I_{t't''}\text{¢}_{ts''}I_{s's}F^{\dagger}(\vec{r}'-\vec{r}'')G(\vec{r}''-\vec{r})F(\vec{r}'-\vec{r})-I_{t't''}\text{¢}_{ts'}I_{ss''}F^{\dagger}(\vec{r}'-\vec{r}'')G(\vec{r}'-\vec{r})F(\vec{r}-\vec{r}'') \\
&+I_{t't''}\text{¢}_{ts}I_{s's''}F^{\dagger}(\vec{r}'-\vec{r}'')G(\vec{r}-\vec{r})F(\vec{r}'-\vec{r}'')-I_{tt''}\text{¢}_{t's}I_{s's''}F^{\dagger}(\vec{r}-\vec{r}'')G(\vec{r}-\vec{r}')F(\vec{r}'-\vec{r}'') \\
&+I_{tt''}\text{¢}_{t's'}I_{ss''}F^{\dagger}(\vec{r}-\vec{r}'')G(\vec{r}'-\vec{r}')F(\vec{r}-\vec{r}'')+I_{tt''}\text{¢}_{t's''}I_{s's}F^{\dagger}(\vec{r}-\vec{r}'')G(\vec{r}''-\vec{r}')F(\vec{r}'-\vec{r}) \\
&+I_{tt'}\text{¢}_{t''s}I_{s's''}F^{\dagger}(\vec{r}-\vec{r}')G(\vec{r}-\vec{r}'')F(\vec{r}'-\vec{r}'')-I_{tt'}\text{¢}_{t''s'}I_{ss''}F^{\dagger}(\vec{r}-\vec{r}')G(\vec{r}'-\vec{r}'')F(\vec{r}-\vec{r}'') \\
&-I_{tt'}\text{¢}_{t''s''}I_{s's}F^{\dagger}(\vec{r}-\vec{r}')G(\vec{r}''-\vec{r}'')F(\vec{r}'-\vec{r})
\end{aligned}
\qquad (29)
$$

It is seen that for an interacting case with the existing of the pairing of two electrons, namely a superconductor, Eq. (29) have the 9 terms more than with respect to non-interacting case due to the combination of normal and anomalous Green's functions. By the definition, for example, $G\left(\vec{r}''-\vec{r}\right)/G\left(0\right)\equiv g_{31}$ and $F\left(\vec{r}''-\vec{r}\right)/G\left(0\right)\equiv f_{31}$ and using $F^{\dagger}\equiv -F$, we can write

$$
\begin{aligned}
&\rho_3 \equiv \rho\left(s,s',s'',t,t',t''\right) = G^3\left(0\right)\\
&\left(+\delta_{t''s}\delta_{t's'}\delta_{ts''}g_{13}g_{31} - \delta_{t''s}\delta_{t's''}\delta_{ts'}g_{13}g_{32}g_{21} - \delta_{t''s'}\delta_{t's}\delta_{ts''}g_{23}g_{12}g_{31} + \delta_{t''s'}\delta_{t's''}\delta_{ts}g_{23}g_{32}\right.\\
&+\delta_{t''s''}\delta_{t's}\delta_{ts'}g_{12}g_{21} - \delta_{t''s''}\delta_{t's'}\delta_{ts} + I_{t't''}\delta_{ts''}I_{s's}f_{23}g_{31}f_{21} + I_{t't''}\delta_{ts'}I_{ss''}f_{23}g_{21}f_{13}\\
&-I_{t't''}\delta_{ts}I_{s's''}f_{23}f_{23} + I_{tt''}\delta_{t's}I_{s's''}f_{13}g_{12}f_{23} - I_{tt''}\delta_{t's'}I_{ss''}f_{13}f_{13} - I_{tt''}\delta_{t's''}I_{s's}f_{13}g_{32}f_{21}\\
&\left.-I_{tt'}\delta_{t''s}I_{s's''}f_{12}g_{13}f_{23} + I_{tt'}\delta_{t''s'}I_{ss''}f_{12}g_{23}f_{13} + I_{tt'}\delta_{t''s''}I_{s's}f_{12}f_{21}\right)
\end{aligned}
\qquad (30)
$$

It should be noted that in general in interacting Fermi gas such as a superconductor, for example $G\left(\vec{r}''-\vec{r}\right)$ is not equal to $G\left(\vec{r}-\vec{r}''\right)$. For d-wave case, we have $G\left(\vec{r}-\vec{r}''\right)=G^{*}\left(\vec{r}''-\vec{r}\right)$ or $g_{13}=g_{31}^{*}$ and thereby, density matrix becomes Hermitian. By considering the following tensor product

$$
\begin{pmatrix}\uparrow\uparrow & \uparrow\downarrow\\ \downarrow\uparrow & \downarrow\downarrow\end{pmatrix}\otimes\begin{pmatrix}\uparrow\uparrow & \uparrow\downarrow\\ \downarrow\uparrow & \downarrow\downarrow\end{pmatrix}\otimes\begin{pmatrix}\uparrow\uparrow & \uparrow\downarrow\\ \downarrow\uparrow & \downarrow\downarrow\end{pmatrix} \qquad (31)
$$

and using the set $\{sts't's''t''\}$ as the labels of the elements of above matrix, the appeared Kronecker delta function and appeared *I* in the elements of reduced density matrix (Eq. (30)) can be obtained. It should be noted that for the d-wave superconductor as s-wave superconductor, we can ignore the anomalous Green's function, *F*. Moreover, three-spin reduced density matrix is given by [29,50]

$$
\begin{aligned}
\rho_3 &= \left(1-p_{12}-p_{13}-p_{23}\right)\frac{I}{8} + p_{12}\left|\Psi_{12}^{-}\right\rangle\left\langle\Psi_{12}^{-}\right|\otimes\frac{I}{2} + p_{13}\left|\Psi_{13}^{-}\right\rangle\left\langle\Psi_{13}^{-}\right|\otimes\frac{I}{2} + p_{23}\left|\Psi_{23}^{-}\right\rangle\left\langle\Psi_{23}^{-}\right|\otimes\frac{I}{2}\\
&= \frac{1}{8}I - \frac{p_{12}}{8}\left(\sigma_x\sigma_x I + \sigma_y\sigma_y I + \sigma_z\sigma_z I\right) - \frac{p_{13}}{8}\left(\sigma_x I\sigma_x + \sigma_y I\sigma_y + \sigma_z I\sigma_z\right) - \frac{p_{23}}{8}\left(I\sigma_x\sigma_x + I\sigma_y\sigma_y + I\sigma_z\sigma_z\right)
\end{aligned}
\qquad (32)
$$

For the investigation of TE, we need $p_{ij}$ in terms of Green's functions. By comparing Eqs. (30) and (32), when considering the equivalent of $g_{ij}$ and $g_{ji}$ (such as for s-wave case), we obtain

$$
p_{ij} = -\frac{-g_{ij}^2 + g_{ij}g_{ik}g_{jk} - f_{ij}^2 + f_{ij}f_{ik}g_{jk} + f_{ij}f_{jk}g_{ik} - f_{ik}f_{jk}g_{ij}}{2+\sum_{\substack{i,j\\ i\neq j,\, j>i}} f_{ij}^2 - \sum_{\substack{i,j\\ i\neq j,\, j>i}} g_{ij}^2 + g_{ij}g_{ik}g_{jk} - f_{ij}f_{jk}g_{ik} - f_{ik}f_{jk}g_{ij} - f_{ij}f_{ik}g_{jk}} \qquad (33)
$$

However, for d-wave case, $p_{ij}$ (given by Eq. (33)) is not true; instead we have (and by neglecting *F* due to smallness)

$$p_{ij} = -\frac{-2g_{ij}g_{ji} + g_{ij}g_{ki}g_{jk} + g_{ji}g_{ik}g_{kj}}{-4 + 2g_{ij}g_{ji} + 2g_{ik}g_{ki} + 2g_{jk}g_{kj} - g_{ij}g_{ki}g_{jk} - g_{ji}g_{ik}g_{kj}} = -\frac{-2g_{ij}g_{ij}^{*} + 2\,\mathrm{Re}\left(g_{ij}g_{ik}g_{jk}\right)}{-4 + 2\sum_{\substack{i,j \\ i\neq j,\, j>i}} g_{ij}g_{ji} - 2\,\mathrm{Re}\left(g_{ij}g_{ik}g_{jk}\right)} \tag{34}$$

Now we proceed to calculate the lower bound of the generalized robustness of TE by using the following formula [31, 39, 40]

$$E_{R,\min}\left(\circ\right) = \underset{ijk\in\{123,231,132\}}{Max}\left\{0, \left(3(p_{ij} + p_{jk}) - 1 - \sqrt{5}\right) \Big/ \left(5 + \sqrt{5}\right)\right\} \tag{35}$$

Previously it was used for the investigation of genuine tripartite entanglement in non-interacting Fermi gas [29,31]. By considering three fermions on a straight line as Refs.[29,31], and considering the distance between 1 and 2 at a fixed $k_F r$ (=0.1) and the distance between particles 1 and 3 as $k_F x$, then $E_{R,\min}$ is plotted in terms of $k_F x$ and $\Delta_0$ at a fixed $r$ for d-wave case (Fig. 2) . As seen, $E_{R,\min}$ depends on energy gap magnitude. From three- dimensional plot (Fig. 2 (a)), it is seen that at a fixed $k_F r$ (for example 0.1), for some energy gap magnitude (provided $\Delta_0 \leq 0.012$), and any value of $k_F x$, $E_{R,\min}$ becomes zero, in spite of the existence of nonzero concurrence (and nonzero BE) in nearly below distance $0.05/k_F\left(\equiv r/2\right)$. However, for s-wave case, this situation does not occur; at any different gap magnitude, there is at least a value of $k_F x$ for which $E_{R,\min}$ is nonzero . Therefore, for d-wave case (s-wave case), gap magnitude dependence (independence) of $E_{R,\min}$ is implied. Furthermore, for d-wave (s-wave) case, our numerical calculation shows, when $r$ is about $0.43/k_F$ ($2.6/k_F$ ), $E_{R,\min}$ starts to become zero for any energy gap magnitude and then for $r \geq 0.43/k_F$ ($r \geq 2.6/k_F$ ), $E_{R,\min}$ is zero. This is not coincided to entanglement length. It merits mentioning that for s-wave case, the result of $E_{R,\min}$ is corresponding to the result given in Ref. [31] that obtained for non-interacting Fermi gas, even thought s-wave superconductor is an interacting system. It should be noted that for d-wave case when $k_F r$ goes to lower values, then, the effect of energy gap magnitude becomes lower to change the value of $E_{R,\min}$ (as an example for $k_F r = 0.01$, Fig. 2(b) was depicted ) and even configuration of the curve related to d-wave case approaches to configuration of the curve related to s-wave case.

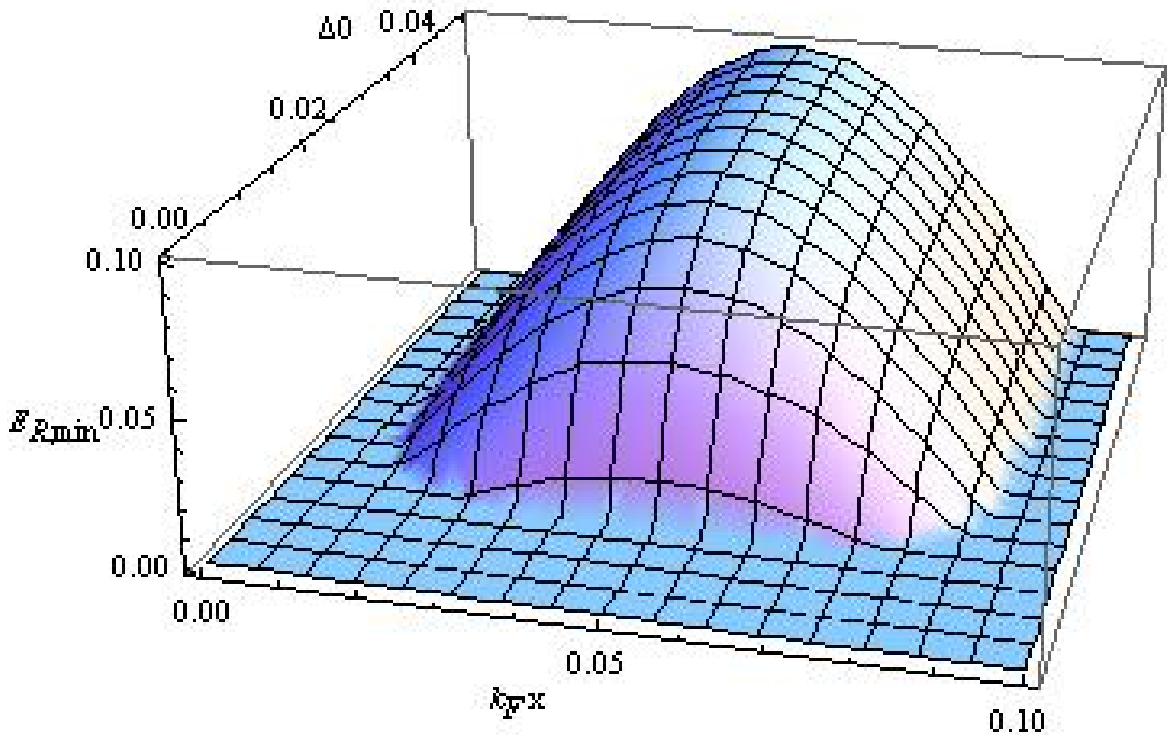

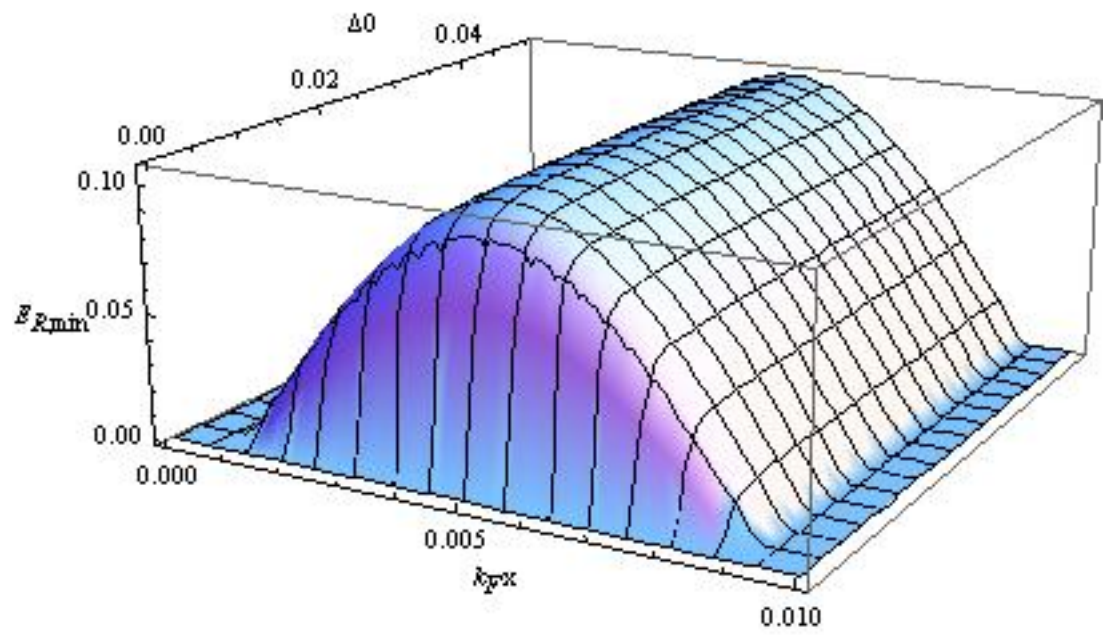

Fig. 2 (a): (Color online) For d-wave case and three fermions on a straight line case with $k_F r = 0.1$.Curve $E_{R,\min}$ in terms of $k_F x$ and $\Delta_0$.

Fig 2(b) ): (Color online) For d-wave case and three fermions on a straight line case with $k_F r = 0.01$. Curve $E_{R,\min}$ in terms of $k_F x$ and $\Delta_0$.

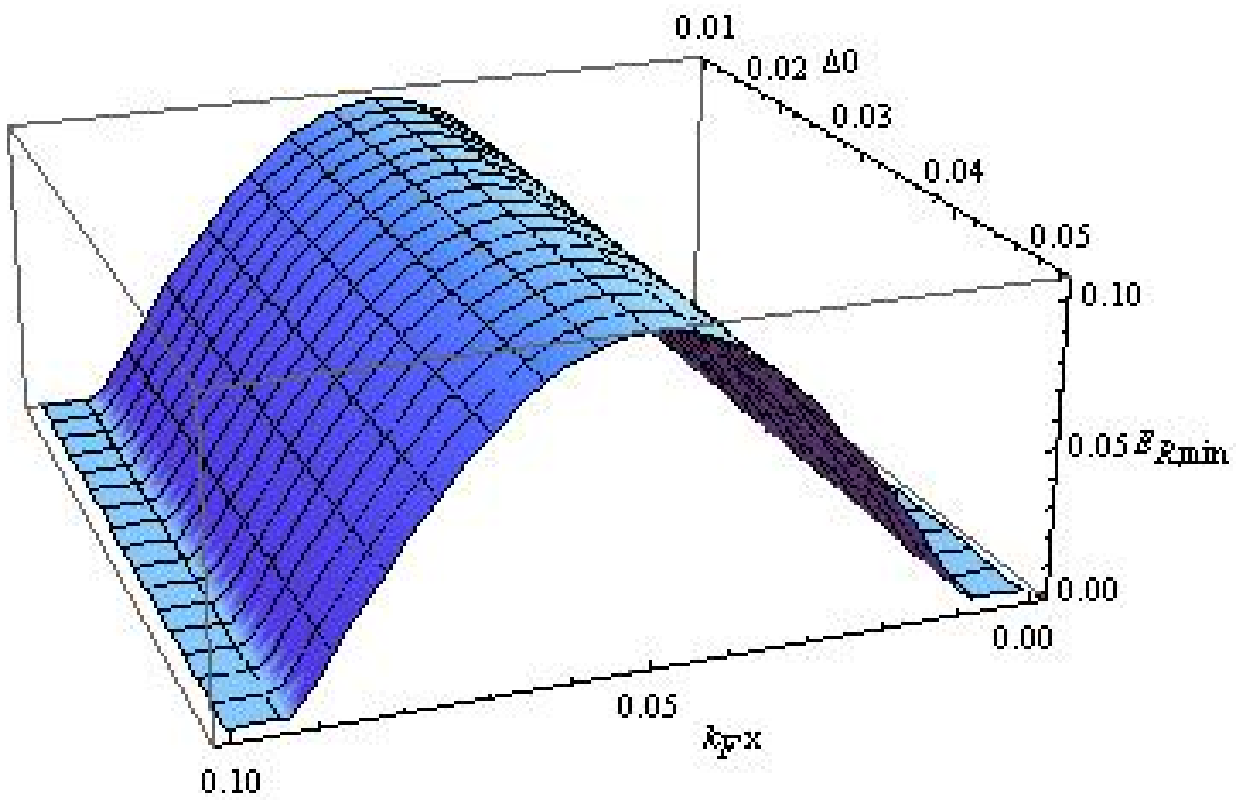

Fig. 3: (Color online) For s-wave case and three fermions on a straight line case with $k_F r = 0.1$. $E_{R,\min}$ in terms of $k_F x$ and $\Delta_0$.

Now we proceed to calculate tripartite in another case (so-call normal case in this paper). As Refs. [29,31], we consider the line that fermion 1 and fermion 3 ( $r$ ) are connected and from the midpoint of the line and normal to it fermion 2 is moved away. $E_{R,\min}$ versus $k_F x$ ( $x$ is normal distance of fermion 2 from the midpoint of $r$ ) and $\Delta_0$ is plotted in Figs. 4 and 5 for d- and s-wave cases, respectively. All results of three fermions on a straight line case are satisfied for the normal case; for example, when $k_F r \simeq 2.6$, then $E_{R,\min} \to 0$ for all the values of gap magnitude. Of course, for normal case, the functional dependence of $E_{R,\min}$ with respect to $k_F x$ is different from the straight line case.

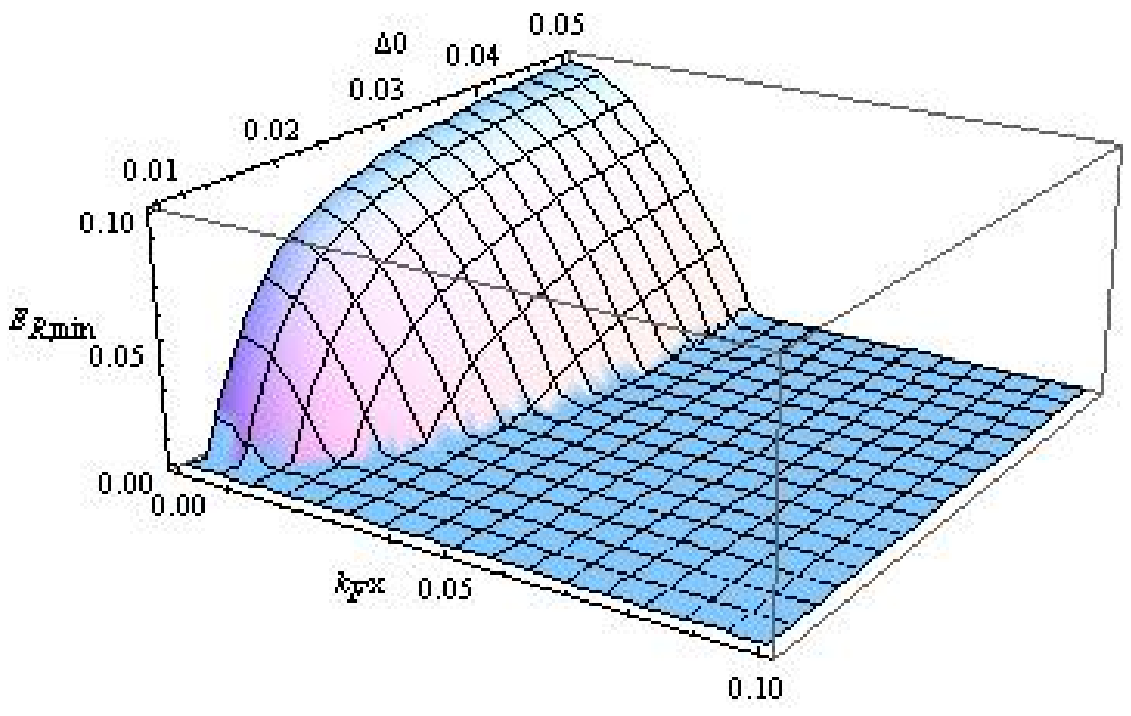

Fig. 4 : (Color online) For d-wave case and normal case with $k_F r = 0.1$. $E_{R,\min}$ in terms of $k_F x$ and $\Delta_0$ .

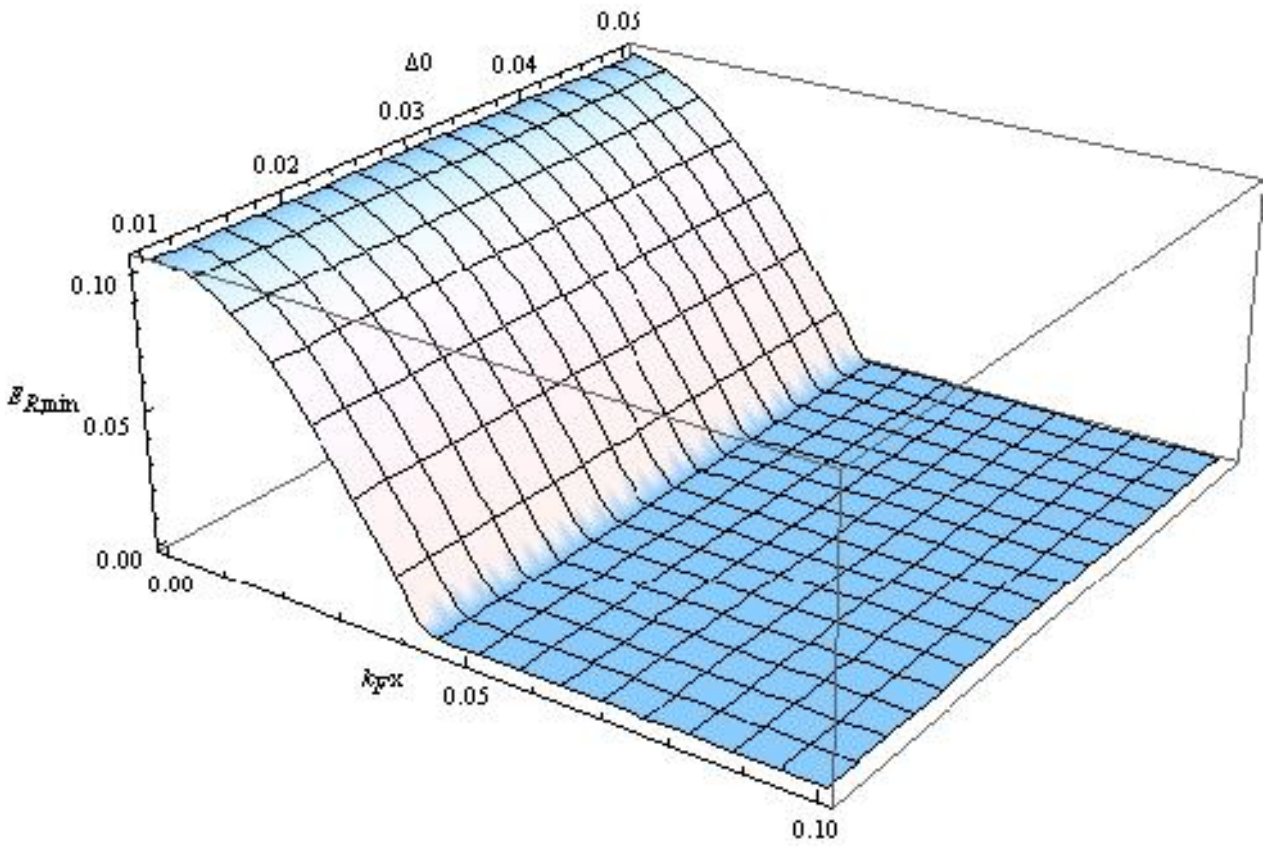

Fig. 5: (Color online) For s-wave case and normal case with $k_F r = 0.1$. $E_{R,\min}$ in terms of $k_F x$ and $\Delta_0$ .

## 4-Quantum discord

QD is mathematically defined as [32-38]

$$Q(\rho) := I(\rho) - C(\rho) \qquad (36)$$

where $I(\rho)$ is quantum mutual information and $C(\rho)$ is classical corrections. $I(\rho)$ is given by

$$I(\rho) = S(\rho^A) + S(\rho^B) - S(\rho^{AB}), \qquad S(\rho) = -tr(\rho \log_2 \rho) \qquad (37)$$

where $S(\rho)$ is so-called von Neumann entropy. For X-states, $I(\rho)$ becomes

$$I(\rho) = S(\rho^A) + S(\rho^B) + \sum_{j=0}^{3} \lambda_j \log_2 \lambda_j \qquad (38)$$

where $\lambda_j$ is the eigenvalue of density matrix, and is given by[37]

$$\lambda_{0,1} = \frac{1}{2}\left[(\rho_{11} + \rho_{44}) \pm \sqrt{(\rho_{11} - \rho_{44})^2 + 4|\rho_{14}|^2}\right], \quad \lambda_{2,3} = \frac{1}{2}\left[(\rho_{22} + \rho_{33}) \pm \sqrt{(\rho_{22} - \rho_{33})^2 + 4|\rho_{23}|^2}\right] \qquad (39)$$

Or the eigenvalues in terms of Green's Function of system under investigation are

$$\lambda = \frac{1}{2}\left(\frac{1 + ig(r)g(-r)}{2 - g(r)g(-r)}\right) \qquad (40)$$

where $\lambda_{0,1,2}$ and $\lambda_3$ are given with $i = -1$ and $i = 1$, respectively. Also, $S(\rho^A)$ and $S(\rho^B)$ are given by

$$S\left(\rho^{A}\right)\left(=-\left[\left(\rho_{11}+\rho_{22}\right)\log_2\left(\rho_{11}+\rho_{22}\right)+\left(\rho_{33}+\rho_{44}\right)\log_2\left(\rho_{33}+\rho_{44}\right)\right]\right)=1 \tag{41}$$

$$S\left(\rho^{B}\right)=\left(S\left(\rho^{A}\right)\text{ with }\rho_{22}\rightleftarrows\rho_{33}\right)=1 \tag{42}$$

Then using Eqs. (38) and (40)-(42), for our system under investigation, $I\left(\rho\right)$ is (with a little different with non-interacting Fermi system)

$$I\left(\rho\right)=1-\log\left(\frac{1-g\left(r\right)g\left(-r\right)}{2-g\left(r\right)g\left(-r\right)}\right)-\frac{1+g\left(r\right)g\left(-r\right)}{2\left(2-g\left(r\right)g\left(-r\right)\right)}\log\frac{1-g\left(r\right)g\left(-r\right)}{1+g\left(r\right)g\left(-r\right)} \tag{43}$$

Furthermore, for the case with the conditions, $\rho_{11}=\rho_{44}$, $\rho_{22}=\rho_{33}$ and real off-diagonal elements of density matrix, $C\left(\rho\right)$ becomes

$$C\left(\rho\right)=S\left(\rho^{A}\right)-\left(-\frac{1+\lambda_j}{2}\log_2\frac{1+\lambda_j}{2}-\frac{1-\lambda_j}{2}\log_2\frac{1-\lambda_j}{2}\right)_{\lambda_j=\max\{\lambda_1,\lambda_2,\lambda_3\}} \tag{44}$$

with

$$\lambda_{1,2}\left(=2\left|\rho_{14}\pm\rho_{23}\right|\right),\quad \lambda_{3,4}\left(=\left|\left(\rho_{11}+\rho_{44}\right)-\left(\rho_{22}+\rho_{33}\right)\right|\right) \tag{45}$$

For d-wave case, we have

$$\lambda_{1,2,3,4}=\frac{g\left(r\right)g\left(-r\right)}{2-g\left(r\right)g\left(-r\right)} \tag{46}$$

Using Eqs. (41), (44) and (46), $C\left(\rho\right)$ becomes

$$C\left(\rho\right)=\log\left(\frac{2}{2-g\left(r\right)g\left(-r\right)}\right)-\left(\frac{1-g\left(r\right)g\left(-r\right)}{-2+g\left(r\right)g\left(-r\right)}\right)\log\left(\frac{1-g\left(r\right)g\left(-r\right)}{-2+g\left(r\right)g\left(-r\right)}\right) \tag{47}$$

Finally, QD can be obtained by using Eqs. (43) and (47) and the numerical calculations of QD is given in Fig. 6

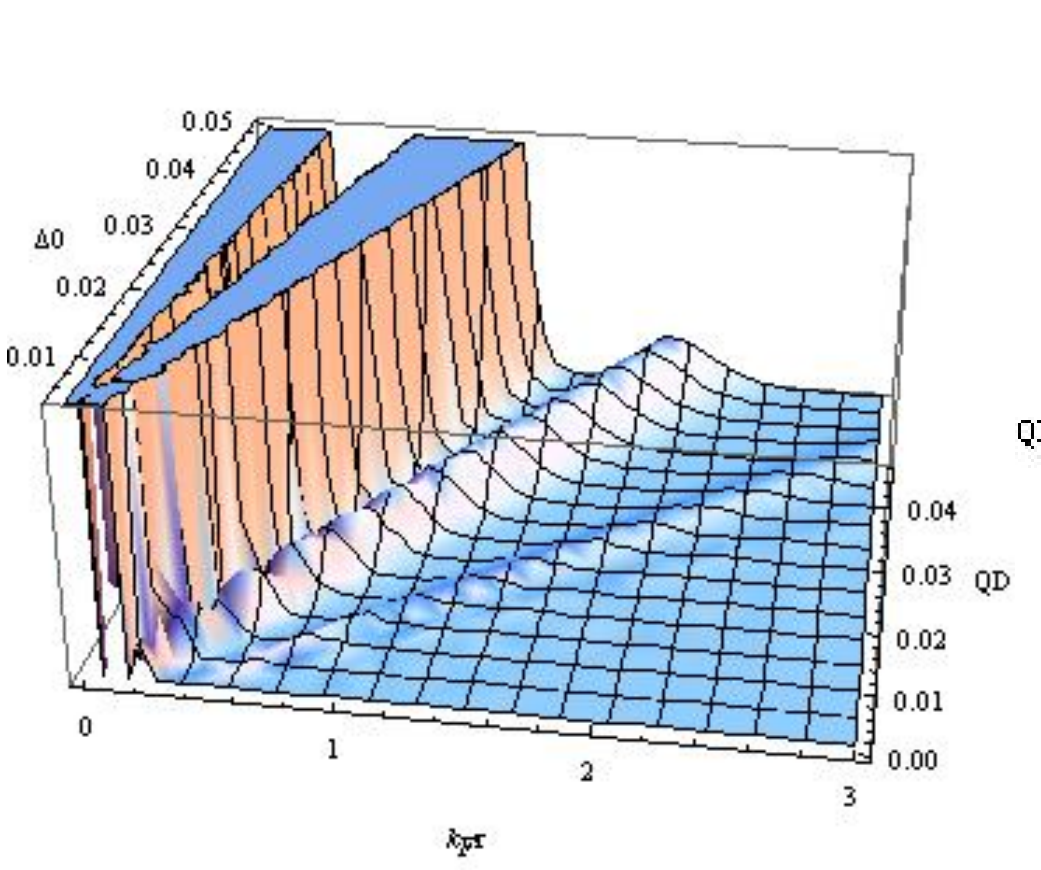

Fig. 6 (a): (Color online) For d-wave case.
QD in terms of $k_F r$ and $\Delta_0$.

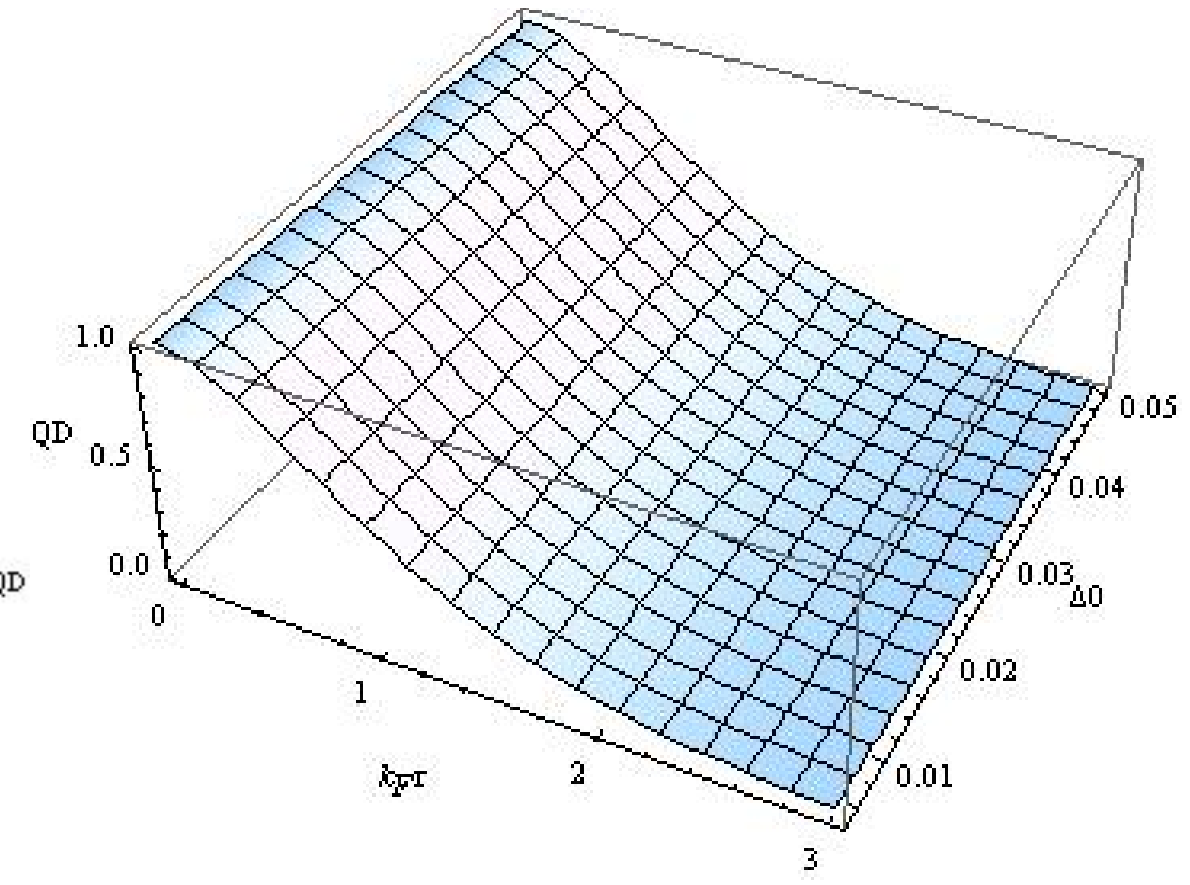

Fig. 6 (b): (Color online) For s-wave case.
QD in terms of $k_F r$ and $\Delta_0$.

In Figs. 6 (a) and 6 (b), QD was plotted versus distance of electron spins of Cooper pair and gap magnitude for d- and s-wave cases, respectively. These figures imply that QD has functional dependence (independence) on gap magnitude for d-wave (s-wave) case. As seen from Figs. 6(a) and 6(c), for d-wave case, QD oscillates with respect to $k_F x$, however for s-wave case (Fig.(b)), it is seen that when the distance of electrons of Cooper pairs increases, quantum correlation i.e. QD reduce (Fig. 6(b)). For d-wave case, whatever the magnitude of gap ($\Delta_0$) becomes higher, effect of gap magnitude on QD goes higher. Movement of quantum correlation length, in which discord becomes zero at all, for two gap magnitude, is given (Fig. 6(c)). When $\Delta_0 = 0.01(0.02)$, quantum correlation length is about $0.275(0.55)$. The slope of curves is different for d- and s-wave cases. Furthermore, from Figs. 1(a) and 6(c), we found that there are the values of $r$ in which BE is zero but QD is nonzero, for example at $r = 0.4/k_F$ .

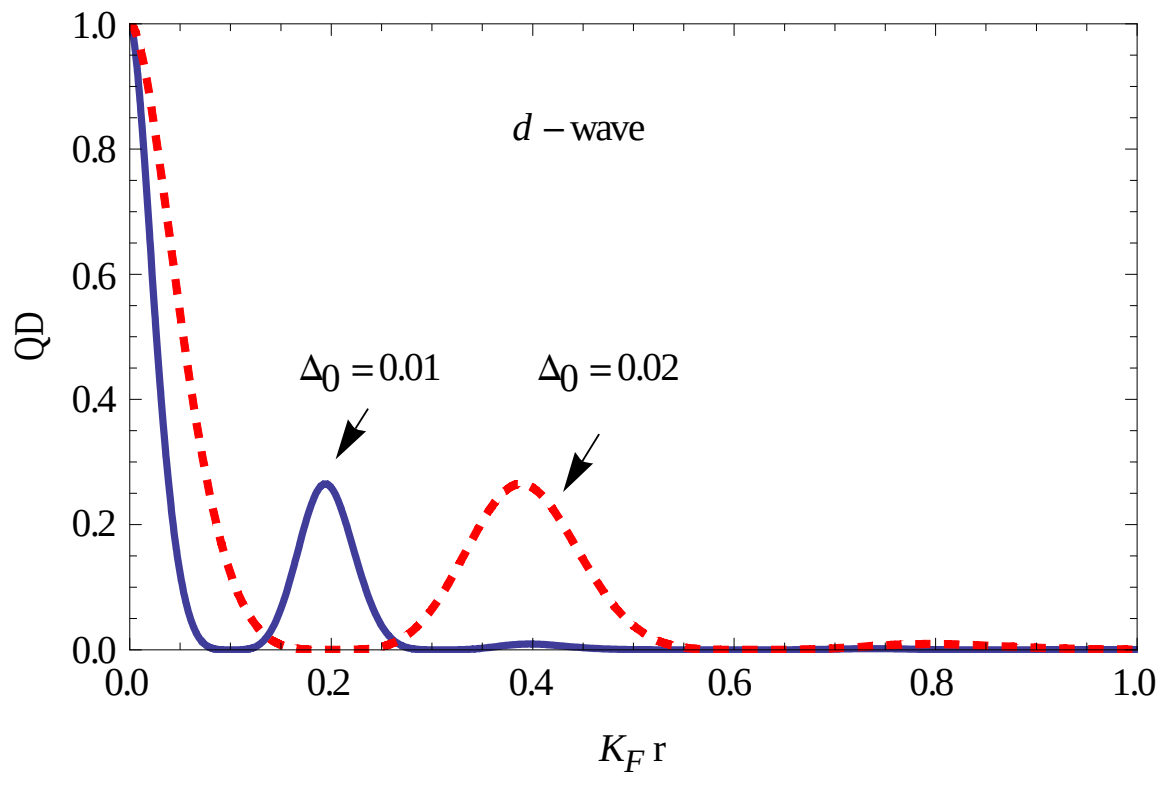

Fig. 6 (c): (Color online) For d-wave case. QD in terms of $k_F r$ at two different $\Delta_0$.

It should be noted that geometric quantum discord, that is given by the following formula [51]

$$D = \frac{1}{2}\sqrt{\frac{¥_1^2 \max\left\{¥_3^2, ¥_2^2 + x_{A3}^2\right\} - ¥_2^2 \min\left\{¥_3^2, ¥_1^2\right\}}{\max\left\{¥_3^2, ¥_2^2 + x_{A3}^2\right\} - \min\left\{¥_3^2, ¥_1^2\right\} + ¥_1^2 - ¥_2^2}} \qquad (48)$$

with the following defined parameters

$$¥_{1,2} = 2(°_{32} \pm °_{41}), \quad ¥_3 = 1 - 2(°_{22} + °_{33}), \quad x_{A3} = 2(°_{11} + °_{22}) - 1 \qquad (49)$$

follows the same obtained results on QD.

## 4. Size effect on system

In this section, we investigate nano-size effect via energy gap fluctuation on the bi- and tri-partite and also discord. Finite-size corrections on BCS superconductivity in metallic nanograins were discussed and semiclassical expansion of spectral density and interaction matrix elements in terms of inverse of system length ($1/k_F L$) were used to obtain the relation between energy gap and the shape and size of system [52]. Now we consider three-dimensional rectangular superconducting grain and the single particle level spacing of it is much smaller than energy gap, which condition ensures to satisfy the BCS theory of superconductivity. Also, the condition $k_F L > 1$ ensures to use semiclassical expansion of spectral density. Then, by considering $\Delta_1$ as gap fluctuation, we have the following transformation to take into account nano-size corrections

$$\Delta_0 \to \Delta' = \Delta_0 + \Delta_1 \qquad (50)$$

where $\Delta_1$ is given by [52]

$$\Delta_1 = \Delta_0\left(f^{(1)} + f^{(3/2)} + f^{(2)}\right) \qquad (51)$$

$f^{(n)}$ is proportional to $\left(1/k_F L\right)^n$. First, in the following we analyze the effect of fluctuation of energy gap on BE via concurrence.

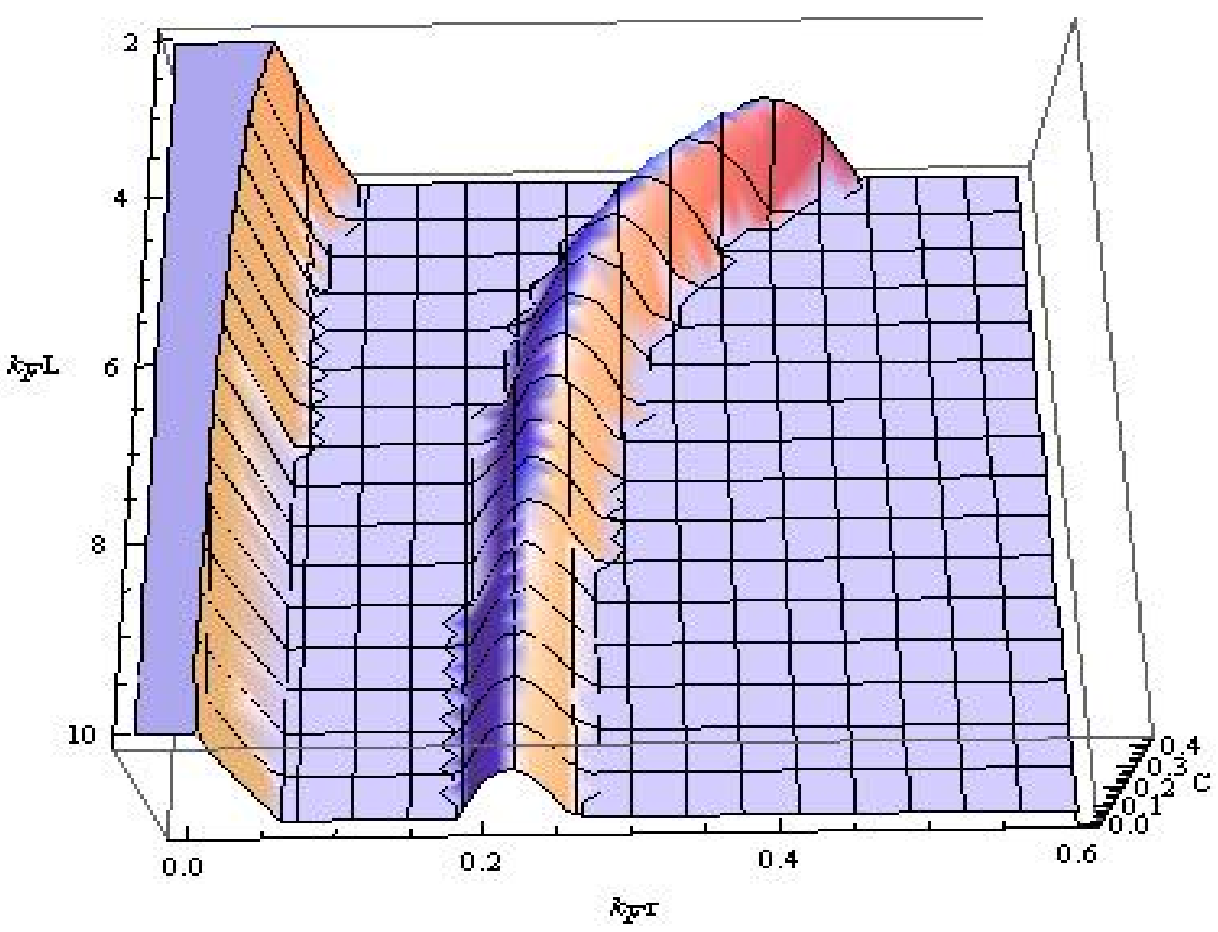

Fig. 7(a): (Color online) For d-wave case with $\Delta_0 = 0.01$. Concurrence in terms of $k_F r$ and $k_F L$.

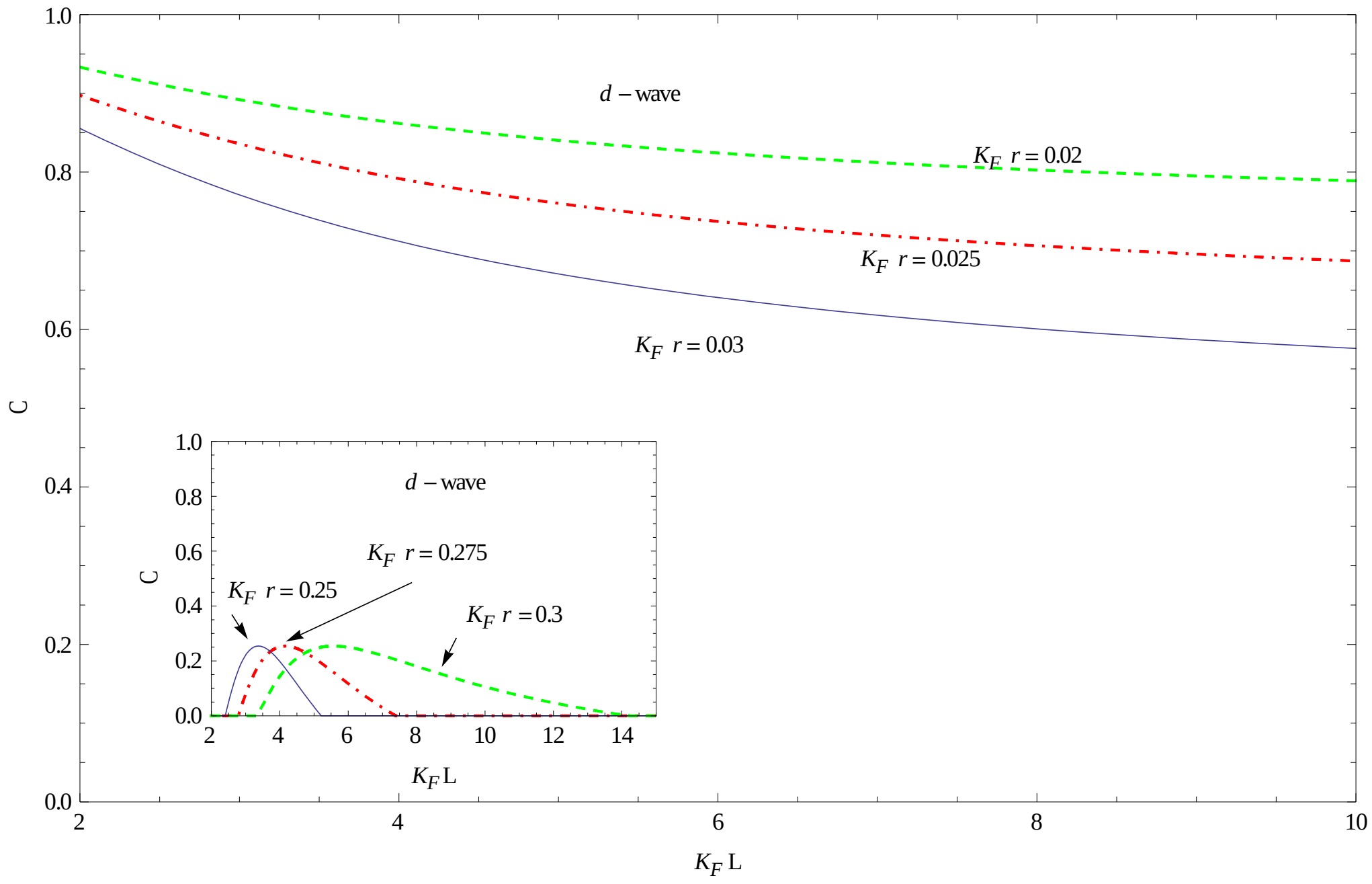

Fig. 7(b): (Color online) For d-wave case with $\Delta_0 = 0.01$. Concurrence in terms of $k_F L$ at different $k_F r$ ( in second principal interval of $k_F r$ ). Inset: Concurrence in terms of $k_F L$ at different $k_F r$ ( in first principal interval of $k_F r$ ).

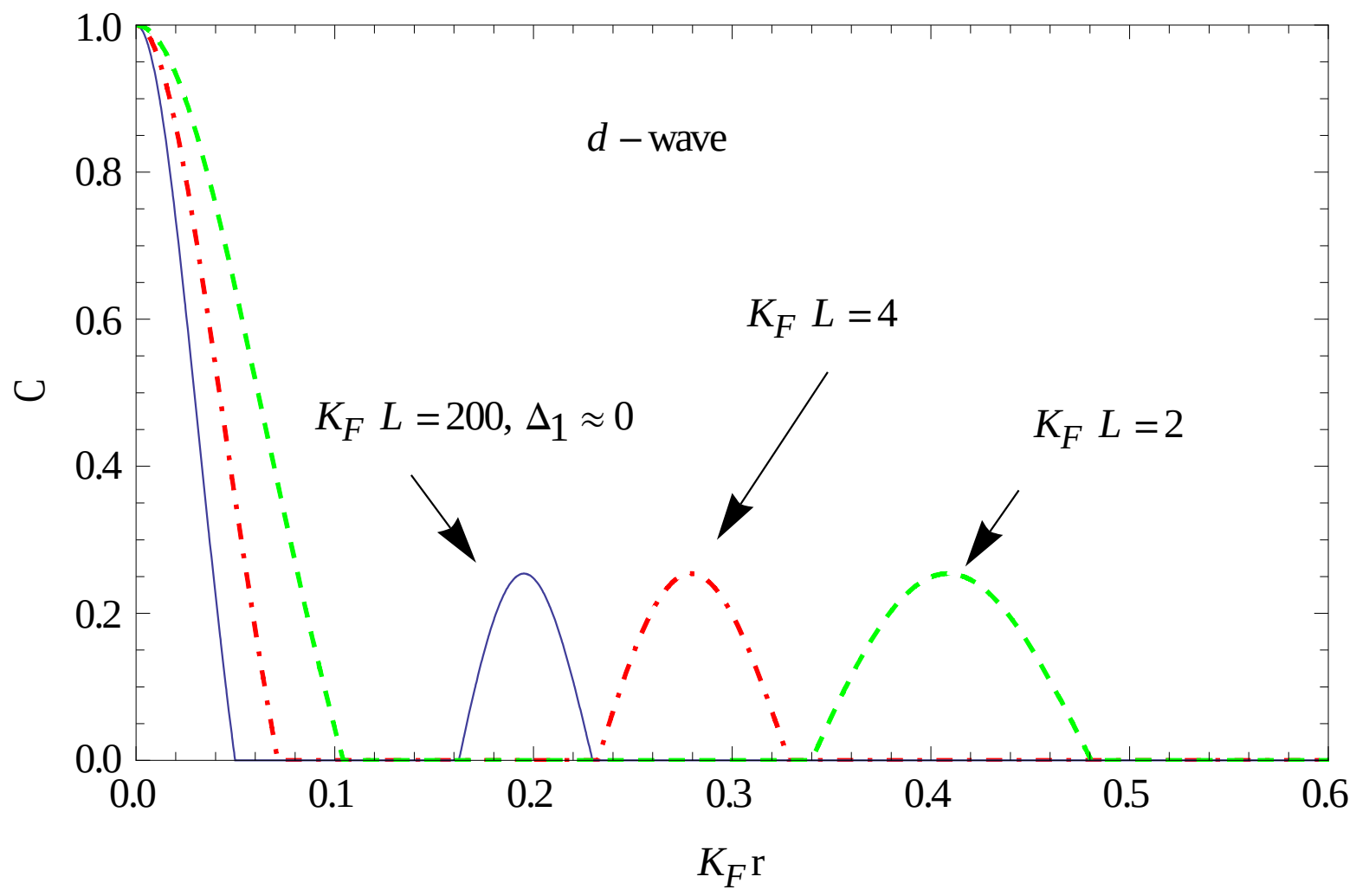

Fig. 7(c): (Color online)For d-wave case and $\Delta_0 = 0.01$. Concurrence in terms of $k_F r$ at different $k_F L$.

Fig. 7 shows the concurrence in terms of the distance of electrons of a Cooper pair for d-wave superconductor (times $k_F$), when considering nano-size effect on system i.e. with existing the fluctuation of energy gap. As seen from Eqs. (50)-(51), by considering nano-size effect, the length of the superconductor($L$) is entered to the concurrence via energy gap. Fig. 7 (a) was plotted versus $k_F r$ and $k_F L$. For d-wave case and by considering $\Delta_0 = 0.01$, concurrence in terms of $k_F L$ at different $k_F r$, in first and second principal interval of $k_F r$, is plotted (Fig. 7(b)) . Functional dependence of concurrence to $k_F L$ is very different in these two intervals. Concurrence in terms of $k_F r$ at different $k_F L$ is depicted in Fig. 7(c) that the oscillation of concurrence in terms of $k_F r$ is clear. Here, it was supposed that when $k_F L$ is $200$, gap fluctuation is absence ($\Delta_1 = 0$) because of the very smallness of value $\Delta_1$. As seen from Fig. 7, the role of length of superconductor is more important for d-wave case. When length of superconductor becomes lower, nano-size effect becomes higher; thereby, nonzero concurrence regions moves to the higher $k_F r$ and broadening of the regions increases. In contrasts to d-wave case, for s-wave case, concurrence is not sensitive to $L$; for any $L$, concurrence is the same as when gap fluctuation is zero.

Now we proceed to analyze of gap fluctuation on QD of system.

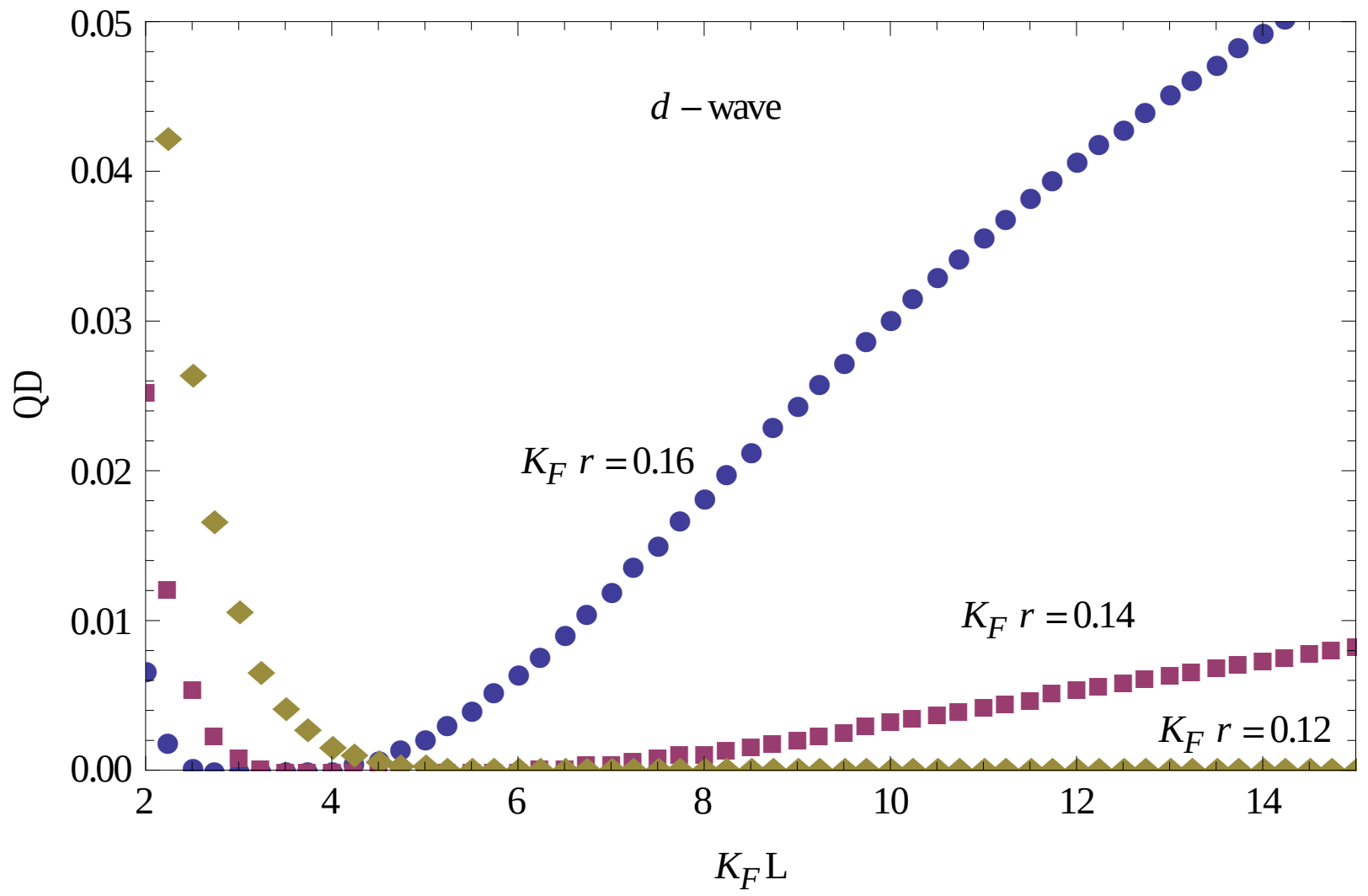

Fig.8(a): (Color online) For d-wave case and $\Delta_0 = 0.01$. QD in terms of $k_F L$ with different $k_F r$ .
The values of $k_F r$ are located in region where the value of concurrence is zero.

For d-wave case, QD versus $k_F L$ at different $k_F r$ was plotted (Fig. 8(a)). The values of $k_F r$ were located in region where the value of concurrence is zero. In spite of having concurrence with zero value, QD has different values. Until now, we didn't interpret this result. Also, we didn't interpret why at any different $k_F r$ , there is a length of superconductor in which QD becomes zero. However, we guess this situation is related to nature of d-wave case.

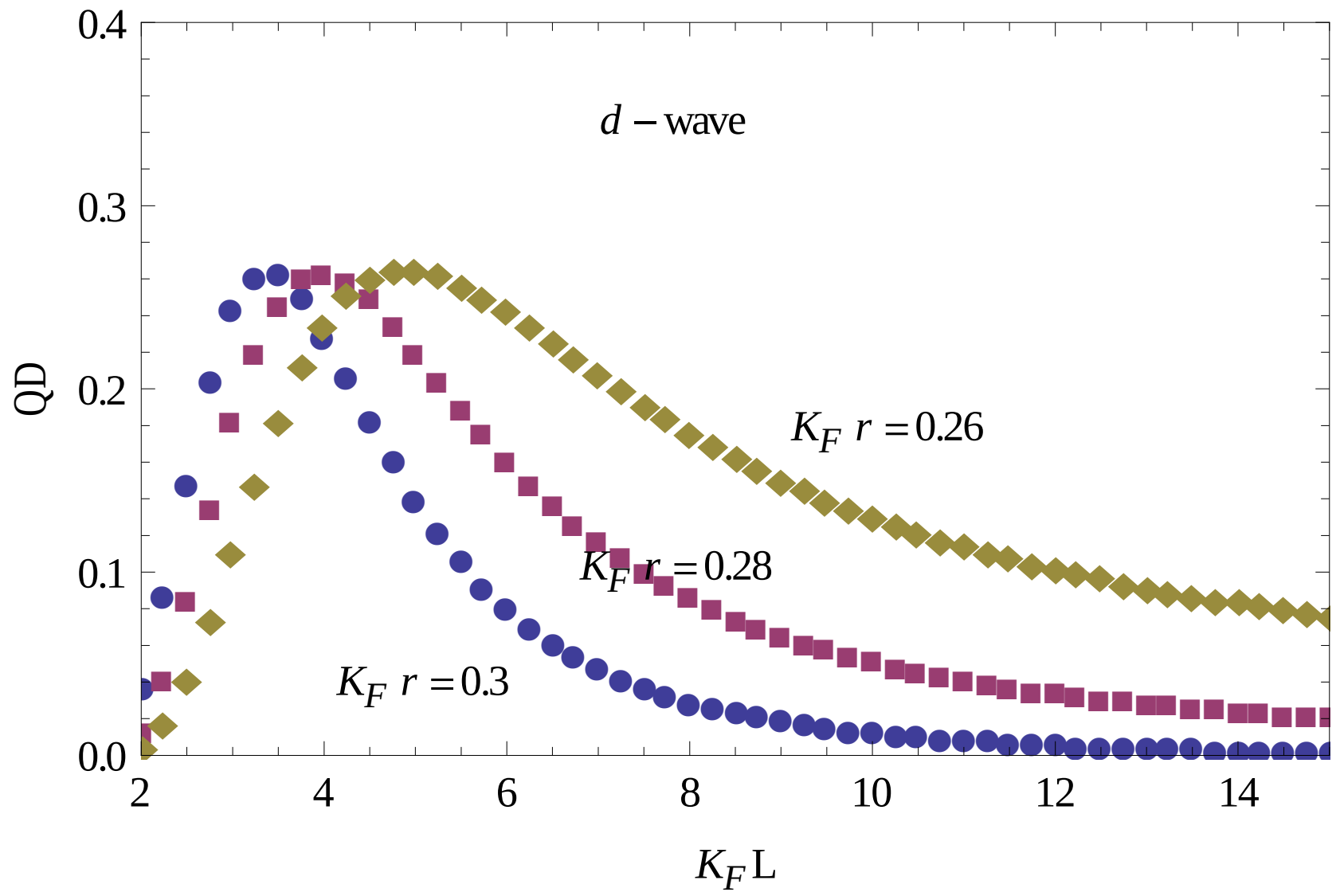

Fig. 8(b): (Color online) For d-wave case with $\Delta_0 = 0.01$. QD in terms of $k_F L$ with different $k_F r$ .
The values of $k_F r$ are located in region where the value of concurrence is nonzero.

In Fig. 8(b), QD in terms of $k_F L$ with different $k_F r$ is plotted. The values of $k_F r$ were located in region where the value of concurrence is nonzero. For a given fixed length of superconductor, both a peak in discord and a peak in concurrence occur simultaneously; for example when $k_F r = 2.8$, peak of QD occur at $k_F L = 4$ (Figs.8 (b) and 8 (c)) and on the other hand, by looking to the curve of concurrence (Fig. 7 (c)) , we find out that peak of concurrence, which occurred at $k_F r = 2.8$, belongs to the curve indicated with $k_F L = 4$. It merits mentioning that the investigation of s-wave case is shown that QD does not depend on length of superconductor and the result is the same as the case without considering nano-size effect.

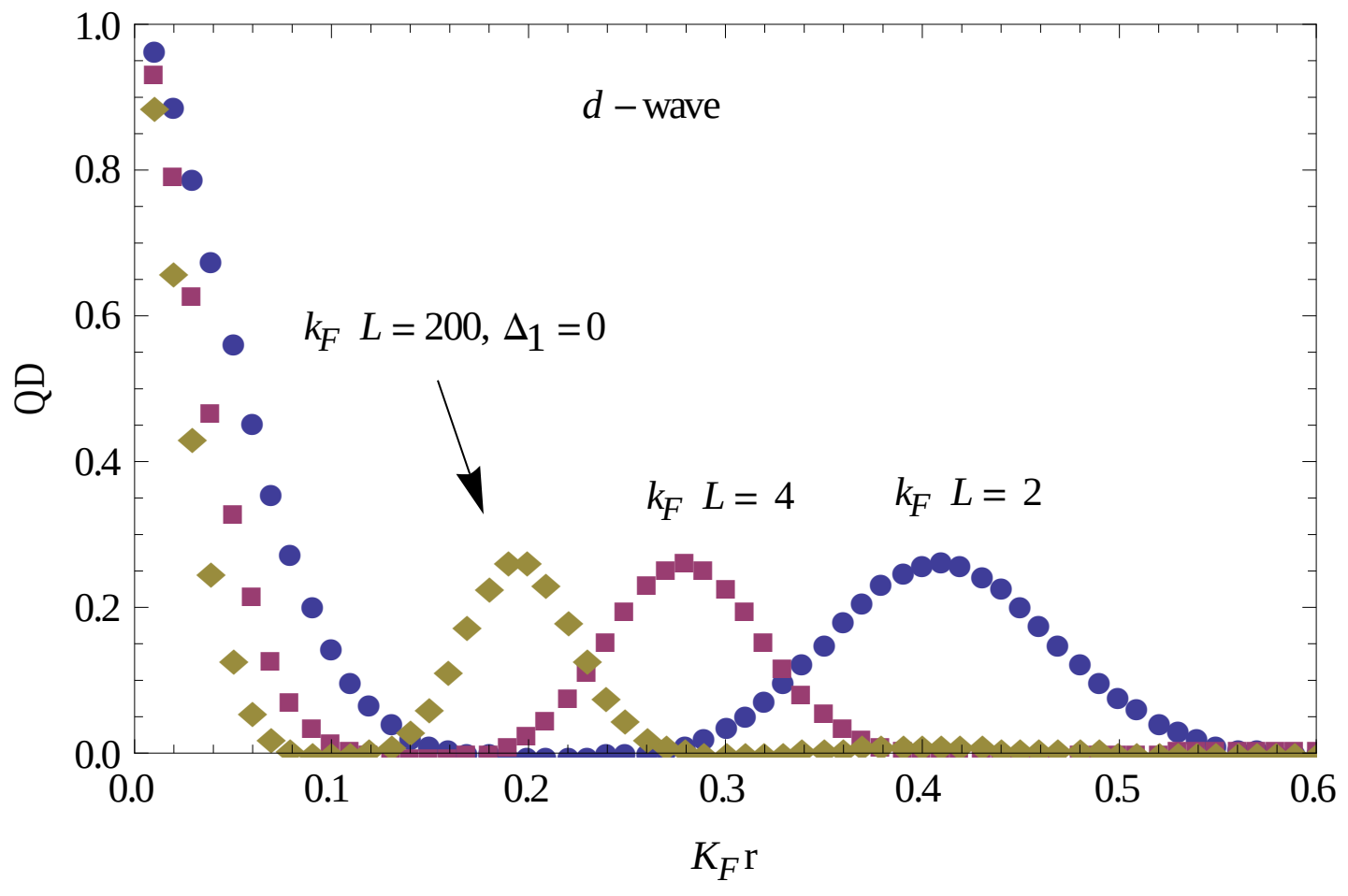

Fig. 8 (c): (Color online) For d-wave case and $\Delta_0 = 0.01$. QD versus $k_F r$ at different $k_F L$.

Fig. 8(c) shows QD with respect to $k_F r$ (at fixed $k_F L$) for d-wave case. Furthermore, the curve without considering nano-size effect was brought in Fig. 8(c) to compare with those accompanied with nano-size effect. Similar to the concurrence, at fixed $k_F L$, the increase on broadening of nonzero QD regions occurs. Also, at fixed $k_F L$, shift of quantum correlation length (QCL) occur and QCL gets higher value, when the length of superconductor becomes lower. Moreover, again QD for s-wave case does not sensitive to the length of superconductor. It should be noted that for d-wave case, we have used the expression of Green's function which consists of energy gap (for d-wave case, It should be remembered that the dependence of momentum of energy gap was disappeared because of the integration on momentum and just gap magnitude was remained), in order to entering the fluctuation of gap.

Now we investigate tripartite in terms of distance of electrons of Cooper pair and length of superconductor. Fig. 9 indicates $E_{R,\min}$ in straight line case as before explained with $k_F r = 0.1$. It is seen that for d-wave case, the curve of $E_{R,\min}$ versus $k_F L$ (Fig. 9 (a)) at $x = r/2$ is located upper to the other curves indicated by $x \neq r/2$; at $x = r/2$ (for the case under investigation $x = 0.05/k_F$ ) $k_F L(=7.9)$, in which $E_{R,\min}$ tends to zero, is highest. Also, at fixed $k_F x$, the increase of $L$ is accompanied by the reduction of $E_{R,\min}$. At different $k_F L$, the curves $E_{R,\min}$ versus $k_F x$ (Fig. 9 (b)) show that there are the increase on $E_{R,\min}$ with lower $L$. Therefore, for nano-scale d-wave superconductor, quantum tripartite is better than for bulk d-wave superconductor. Meanwhile, for the case $k_F r = 0.1$, at $k_F L = 200$ (or $\Delta_1 = 0$ namely without gap fluctuation), $E_{R,\min}$ is zero (Fig. 9 (b)) for all range of $k_F x$. However, in contrast to d-wave case, bulk and nano-size s-wave superconductor are the same (Fig. 9 (c)).

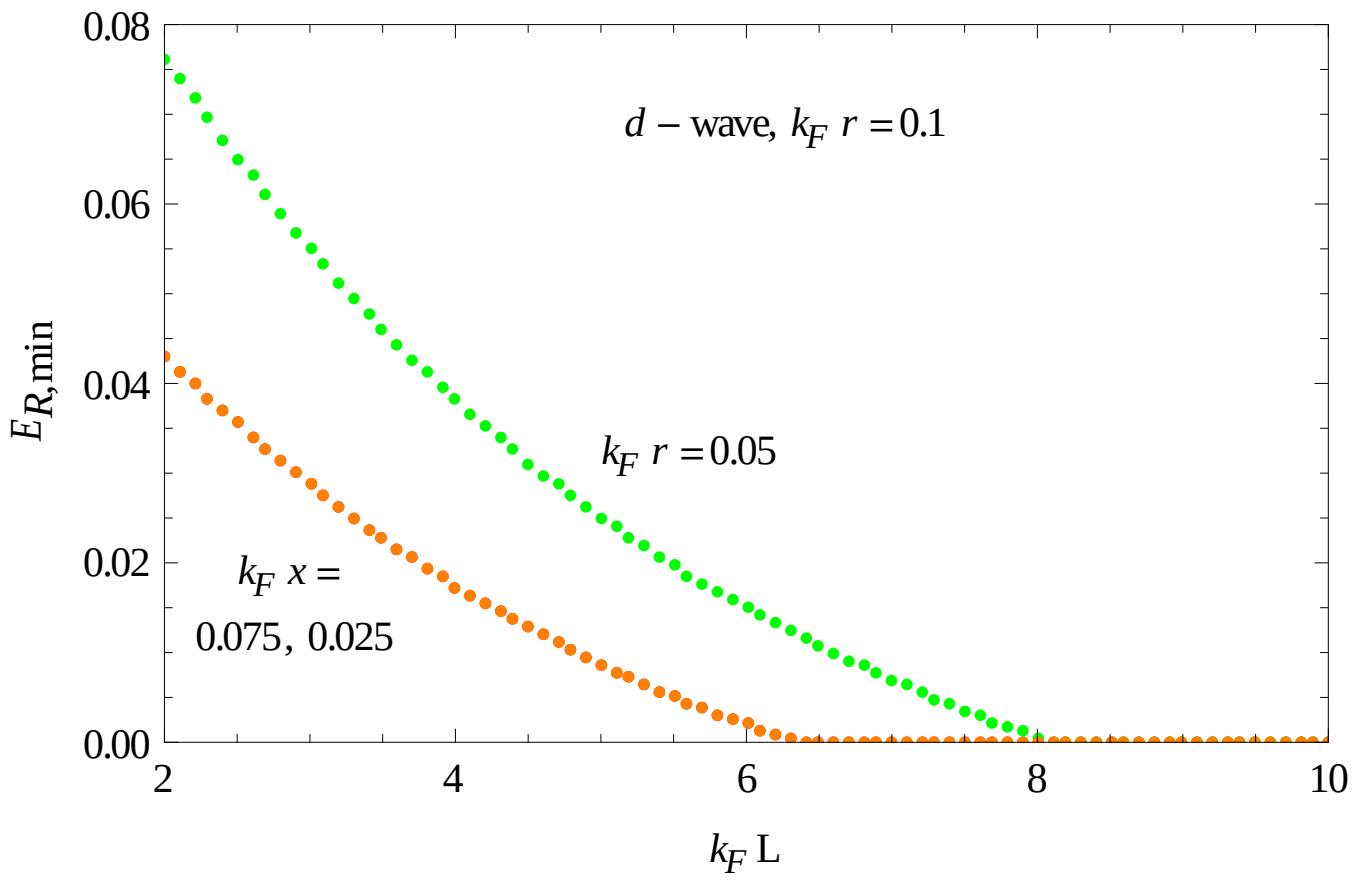

Fig. 9(a): (Color online) For d-wave case in straight line case with $\Delta_0 = 0.01$ and $k_F r = 0.1$ . $E_{R,\min}$ in terms of $k_F L$ with different $k_F x$ .

Also, it should be noted that for d-wave case, $E_{R,\min}$ is negligible for any $k_F L$ and $k_F x$,provided $k_F r > 0.18$. For d-wave case, the role of the nano-size effect becomes more important as the $L$ decreases.

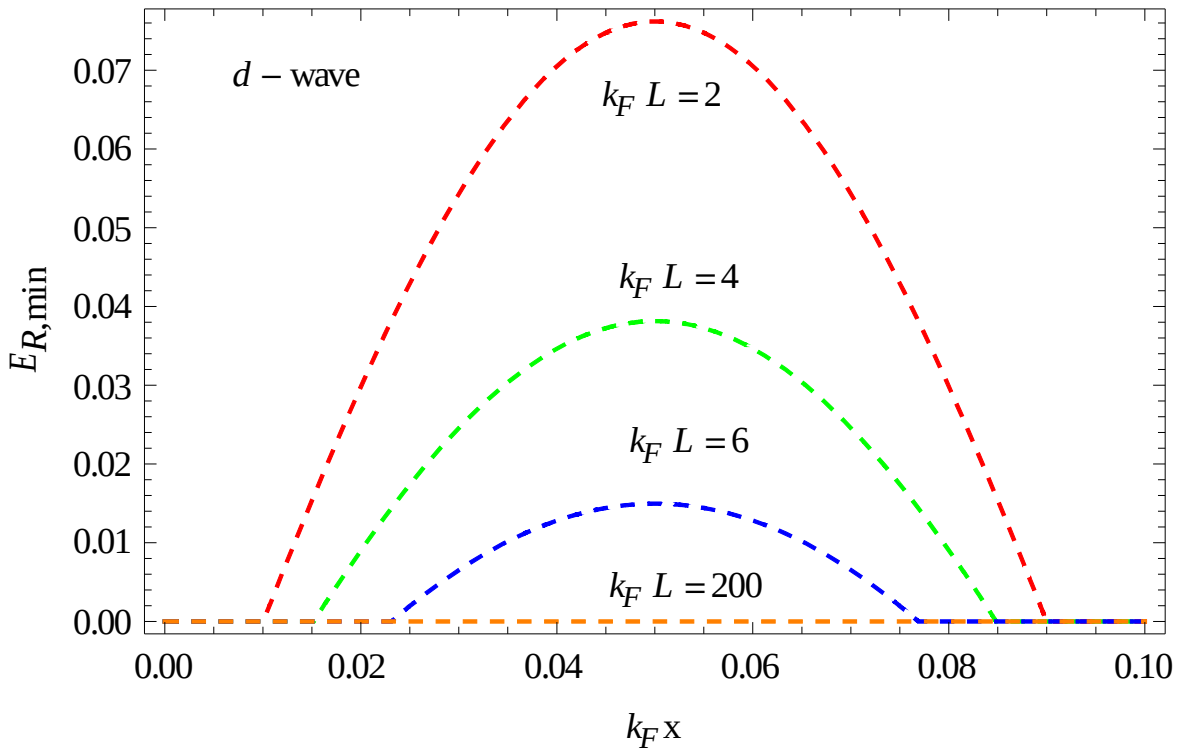

Fig. 9 (b): (Color online) For d-wave case in straight line case with $\Delta_0 = 0.01$ and $k_F r = 0.1$. $E_{R,\min}$ in terms of $k_F x$ with different $k_F L$.

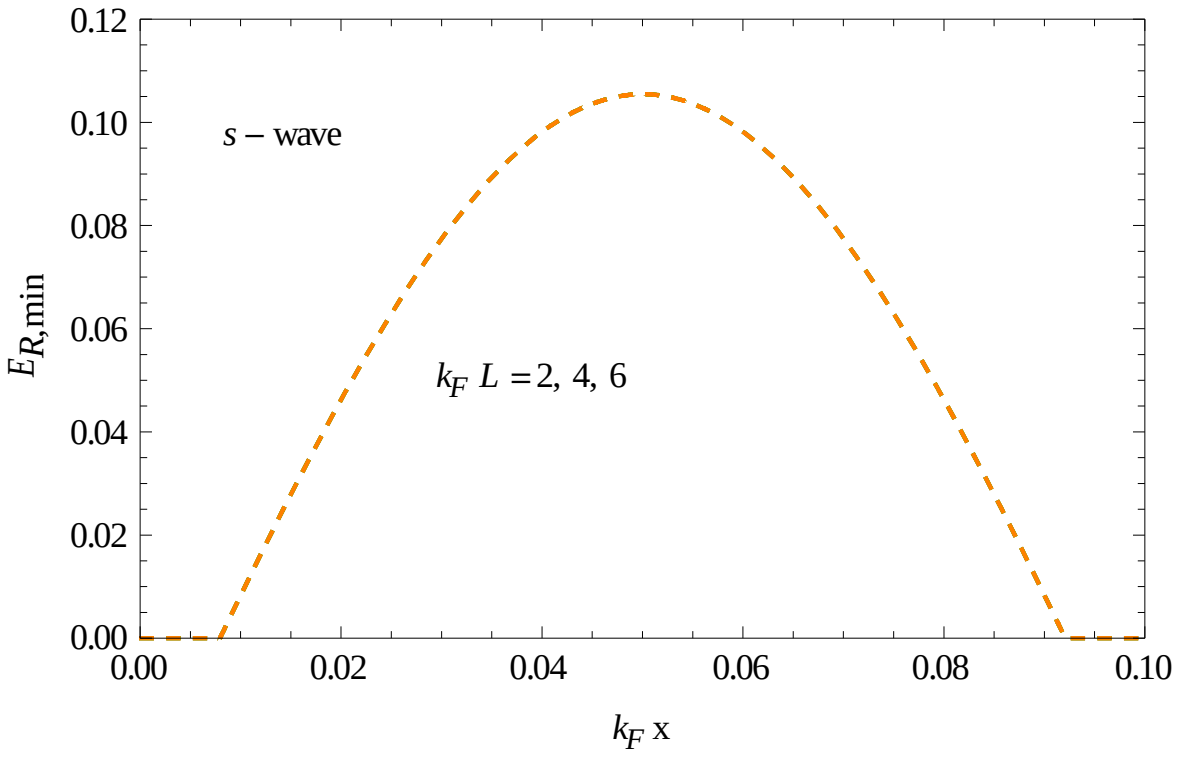

Fig. 9 (c): (Color online) For s-wave case in straight line case with $\Delta_0 = 0.01$ and $k_F r = 0.1$. $E_{R,\min}$ in terms of $k_F x$ with different $k_F L$.

Now we deal with tripartite in normal situation. For d-wave case as shown in Fig. 9(d), at fixed $k_F x$, the nano-size effect accompanied by with the reduction of *L*, causes the increase of $E_{R,\min}$. As shown in Fig. 9(d), there are some values of *L*, in which $E_{R,\min}$ start to have zero value, at each value of $k_F x$. For higher values of fixed $k_F x$, zeroes of $E_{R,\min}$ occur sooner. However, s-wave case does not affect by length of superconductor. For d-wave case, at a fixed $k_F L$, $E_{R,\min}$ is the deceasing function of $k_F x$ (Fig. 9(e)) and also at lower $k_F L$,the value of $k_F x$, in which $E_{R,\min}$ becomes zero, increases.

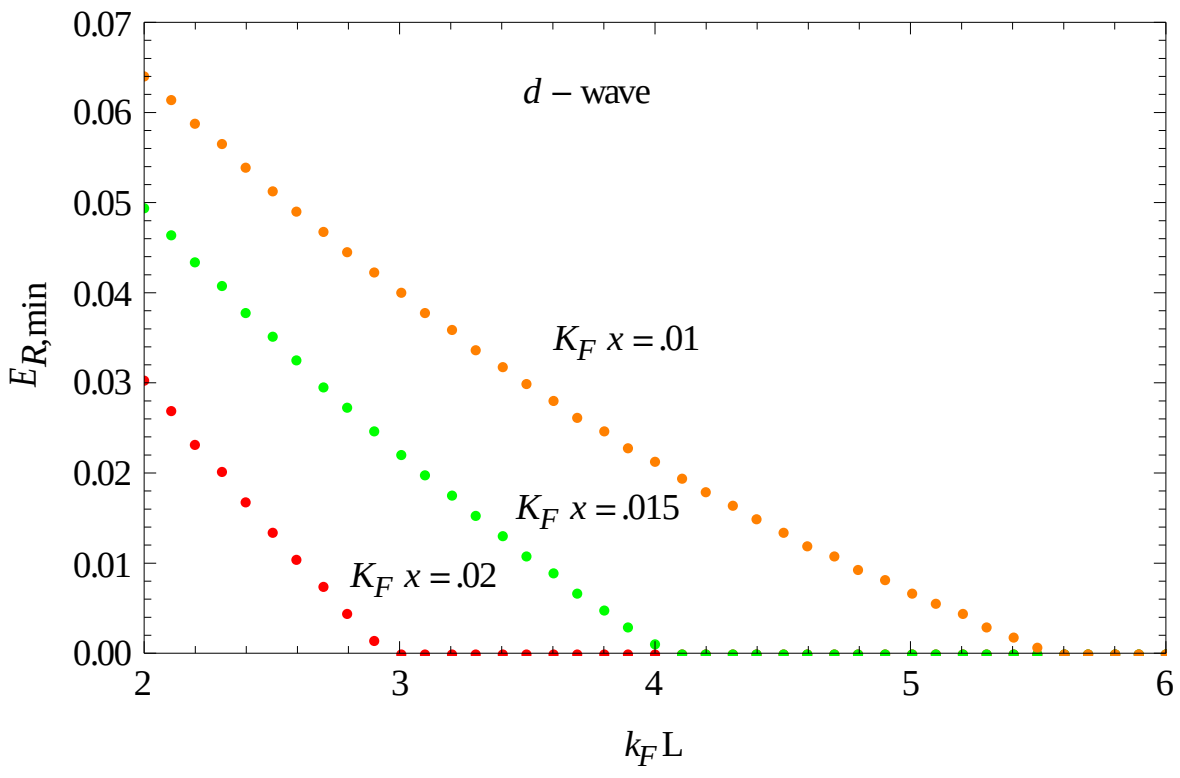

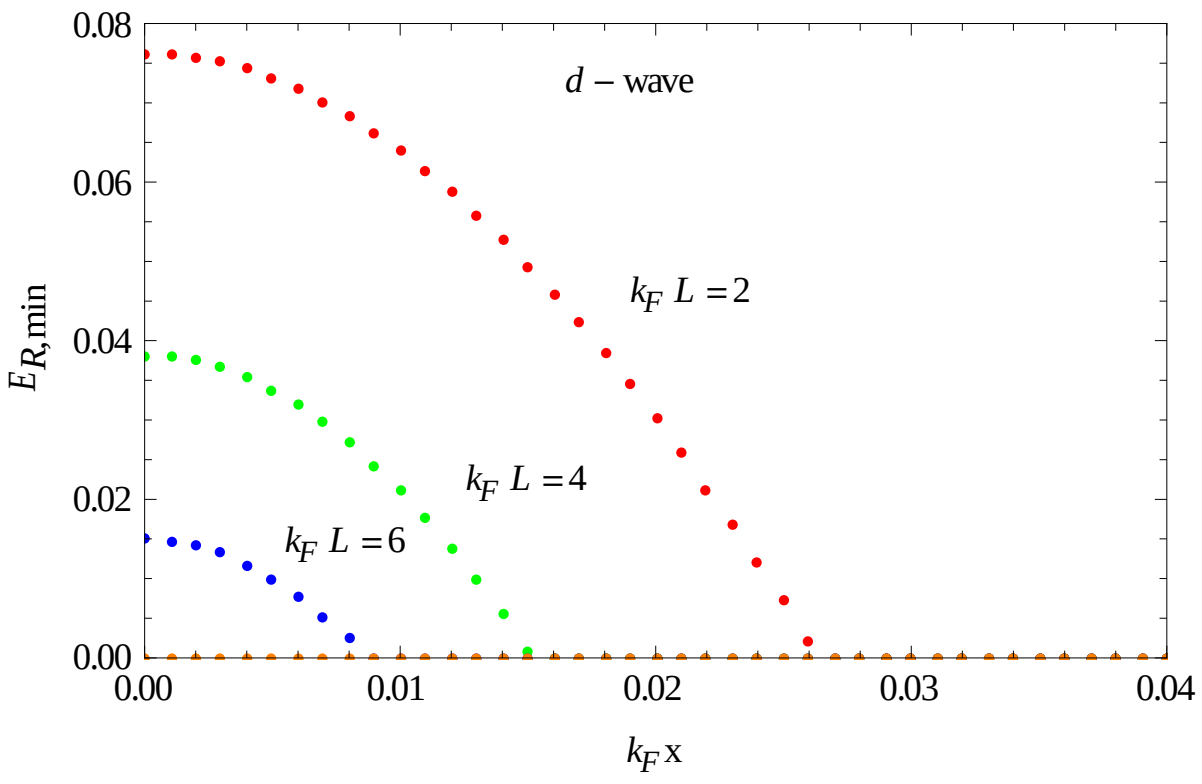

Fig. 9 (d): (Color online) For d-wave case in normal case with $\Delta_0 = 0.01$ and $k_F r = 0.1$. $E_{R,\min}$ in terms of $k_F L$ with different $k_F x$.

Fig. 9 (e): (Color online) For d-wave case in normal line case with $\Delta_0 = 0.01$ and $k_F r = 0.1$. $E_{R,\min}$ in terms of $k_F x$ with different $k_F L$.

Also, for normal case and d-wave (s-wave) case, three-dimensional depict of $E_{R,\min}$ in terms of $k_F x$ and $k_F L$ was brought in Fig. 9 (f) (Fig. 9 (g)) and as a discrepancy of d- and s-wave cases can be said that the existence of the functional dependence of d-wave case to $L$ , in contrast to s-wave.

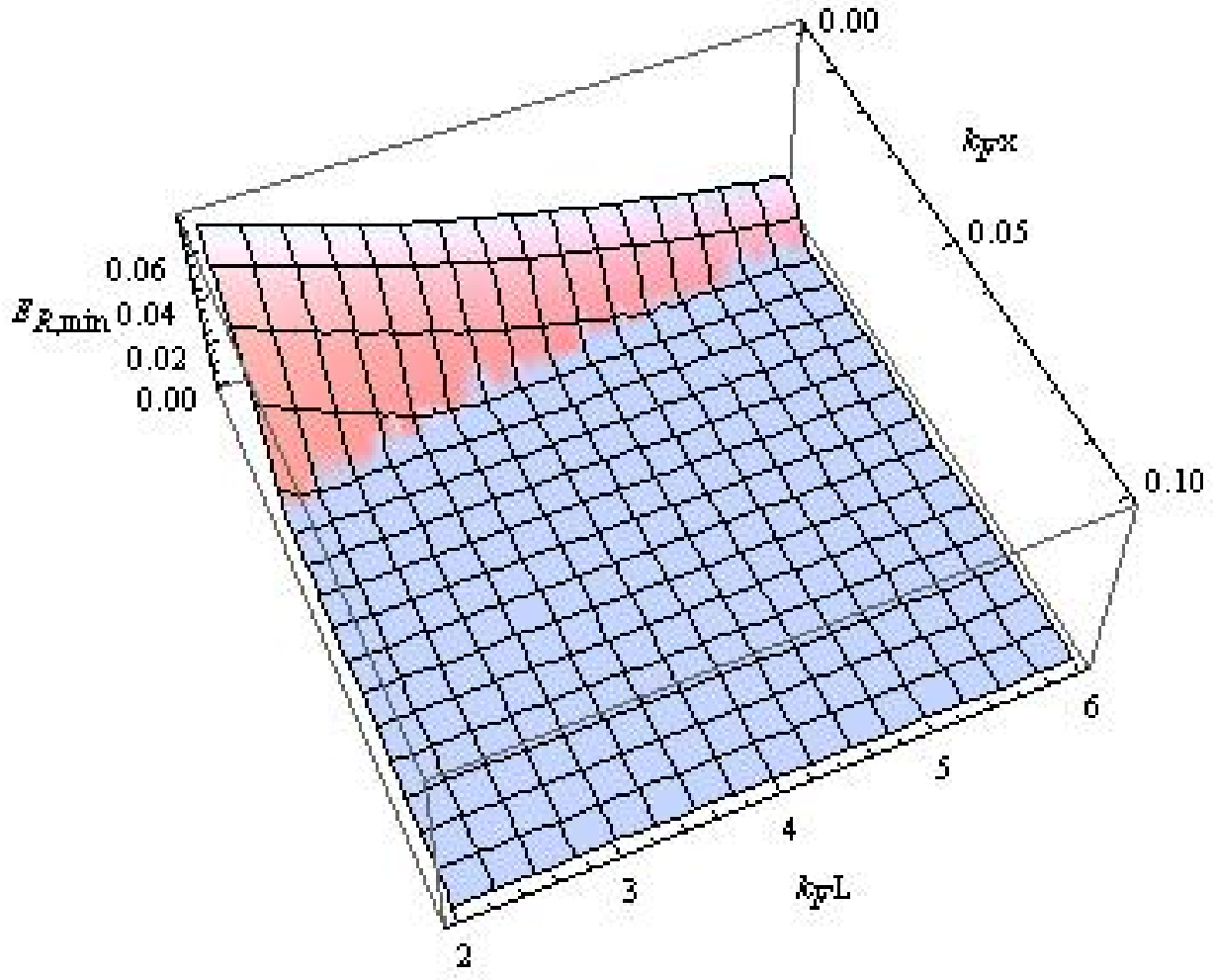

Fig. 9(f): (Color online) For d-wave and normal case with $\Delta_0 = 0.01$ and $k_F r = 0.1$. $E_{R,\min}$ in terms of $k_F x$ and $k_F L$.

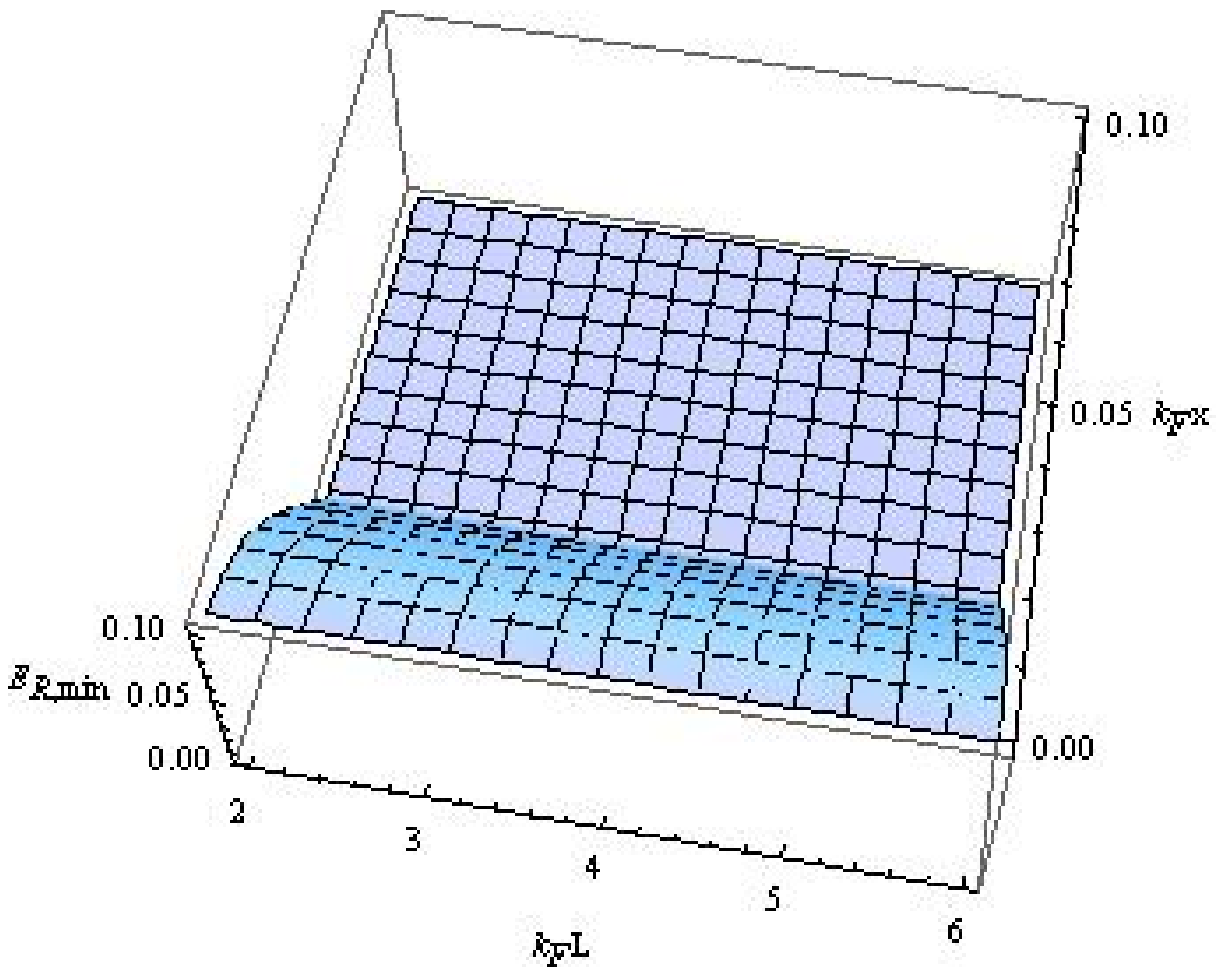

Fig. 9(g): (Color online) For s-wave and normal case with $\Delta_0 = 0.01$ and $k_F r = 0.1$. $E_{R,\min}$ in terms of $k_F x$ and $k_F L$.

## 5. Conclusions

Green's functions (Eqs. (24)-(25)) of an interacting Fermi system namely d-wave superconductor were calculated. The lower bound of the generalized robustness of TE of system with new parameters (Eqs. (33)-(34)) was obtained on base of spin-space density matrix. We investigated quantum correlation via BE, $E_{R,\min}$ and QD of the d-wave superconductor. Moreover, nano-size effect via gap fluctuation was entered. The relation of the quantities concurrence, $E_{R,\min}$ and QD to the energy gap magnitude, the length of superconductor and the distance of two electrons of Cooper pair were obtained. Quantum correlation length and entanglement length and the influence of nano-size effect on system were investigated. The discrepancies of the results for d- and s-wave cases were discussed in details. The existence of nodes in energy gap causes to influence on quantum correlation very well and also the nodes cause the role of size of system on quantities related to quantum correlation becomes more important.